\documentclass[12pt]{iopart}

\usepackage{color, graphicx}
\usepackage{amssymb, amsmath, slashed}
\usepackage{cite}

\begin{document}

\title[Neutrinos and Collider Physics]{Neutrinos and Collider Physics}

\author{Frank~F.~Deppisch$^1$, P. S. Bhupal~Dev$^2$, Apostolos Pilaftsis$^2$}
\address{$^1$\em Department of Physics and Astronomy, University College London, \\ 
London WC1E 6BT, United Kingdom\\
$^2$Consortium for Fundamental Physics, School of Physics and Astronomy, \\
University of Manchester, Manchester M13 9PL, United Kingdom}
\ead{f.deppisch@ucl.ac.uk, bhupal.dev@manchester.ac.uk, apostolos.pilaftsis@manchester.ac.uk}




\begin{abstract}
We review the collider phenomenology of neutrino physics and the synergetic aspects at energy, intensity and cosmic frontiers to test the new physics behind the neutrino mass mechanism. In particular, we focus on seesaw models within the minimal setup as well as with extended gauge and/or Higgs sectors, and on supersymmetric neutrino mass models with seesaw mechanism and with $R$-parity violation. In the simplest Type-I seesaw scenario with sterile neutrinos, we summarize and update the current experimental constraints on the sterile neutrino mass and its mixing with the active neutrinos. We also discuss the future experimental prospects of testing the seesaw mechanism at colliders and in related low-energy searches for rare processes, such as lepton flavor violation and neutrinoless double beta decay.  The implications of the discovery of lepton number violation at the LHC for leptogenesis are also studied. 
\end{abstract}


\section{Introduction} \label{sec:1} 
With the discovery of a Higgs boson at the Large Hadron Collider (LHC)~\cite{Aad:2012tfa, Chatrchyan:2012ufa} having properties consistent with the Standard Model (SM) expectations~\cite{Agashe:2014kda}, we are tantalizingly close to verifying the Higgs mechanism responsible for the SM gauge boson and charged fermion masses. What remains missing though is an understanding of the neutrino masses. The observation of neutrino oscillations in solar, atmospheric, reactor and accelerator neutrino data~\cite{Agashe:2014kda}  demonstrates that at least two of the three active neutrinos have a non-zero mass and that individual lepton flavor is violated. This provides the first and so far {\em only}  conclusive experimental evidence for the existence of New Physics beyond the SM. 

In the SM, neutrinos have only one helicity state $\nu_L$, and therefore, cannot acquire a Dirac mass, unlike the charged fermions, after the electroweak  symmetry breaking (EWSB) by the vacuum expectation value (VEV) of the SM Higgs field. Just adding `by hand' a right-handed (RH) neutrino field $N$ per generation to the SM, one could in principle generate a Dirac mass term for neutrinos.   However, to get sub-eV left-handed (LH) neutrino masses as required by the neutrino oscillation data, one needs the Dirac Yukawa couplings to be extremely tiny $\lesssim 10^{-12}$. There is no theoretical justification for the large disparity between such small neutrino Yukawa couplings and other SM Yukawa couplings. Moreover, such a scenario  would be rather uninteresting from an experimental point of view. Therefore, we will take the more optimistic viewpoint that some other New Physics might be responsible for the observed smallness of neutrino masses.  

Being electrically neutral, neutrinos can in principle have a Majorana mass term of the form $\bar{\nu}_{L\ell}^C\nu_{L\ell}$, where $\nu_L^C\equiv \nu_L^{\sf T}C^{-1}$, $C$ being the charge conjugation matrix, and $\ell=e,\mu,\tau$ is the lepton flavor index. However, since $\nu_{L\ell}$ is part of the $SU(2)_L$ doublet field $L_\ell=(\nu_\ell, \ell)_L^{\sf T}$ with lepton number $L=+1$, the above Majorana mass term transforms as an $SU(2)_L$ triplet, i.e.~it is {\em not} gauge invariant, apart from breaking the global $L$ and $B-L$ symmetries of the SM by two units. On the other hand, the only known source of lepton number violation (LNV) in the SM is via non-perturbative weak instanton effects 
through the Adler-Bell-Jackiw anomaly~\cite{Adler:1969gk, Bell:1969ts}. These non-perturbative effects only violate the $B+L$ symmetry~\cite{tHooft:1976up}, but preserve the $B-L$ symmetry to all orders.  Therefore, neutrino masses cannot be induced even by non-perturbative effects in the SM. 

One might wonder whether quantum gravity effects might be sufficient to explain the tiny neutrino masses. As long as gravity is treated perturbatively and respects the $B-L$ symmetry of the SM, neutrinos remain massless. However, non-perturbative gravitational effects, e.g. black holes and worm holes, do not respect global symmetries and could induce non-zero neutrino masses in the effective low energy Lagrangian. In the context of the SM, these effects are at most of order $v^2/M_{\rm Pl} \sim 10^{-5}$ eV~\cite{Barbieri:1979hc, Akhmedov:1992hh} (where $v$ is the SM Higgs VEV and $M_{\rm Pl}$ is the Planck mass), which is still three orders of magnitude below that required to satisfy the atmospheric neutrino data~\cite{Agashe:2014kda}. 

Therefore, it seems more {\em natural} to invoke some $(B-L)$-violating New Physics beyond the SM at a scale $\Lambda\ll M_{\rm Pl}$ to explain the observed neutrino masses~\cite{Mohapatra:2006gs}.  
Within the SM, these $(B-L)$-breaking effects can be parametrized through an effective dimension-5 Weinberg operator of the form $\lambda_{\ell\ell'}L_\ell L_{\ell'}\Phi \Phi/\Lambda$~\cite{Weinberg:1979sa}, where $\Phi=(\phi^+,\phi^0)^{\sf T}$ is the SM Higgs doublet. It is easy to see that there are {\em only} three ways to obtain this Weinberg operator at tree-level, using only renormalizable interactions, i.e., 
\begin{enumerate}
\item [(i)] The product of $L_\ell$ and $\Phi$ forms a fermion singlet: $(L_\ell^{\sf T}\Phi)(L_{\ell'}^{\sf T}\Phi)/\Lambda$. This is widely known as the {\em Type-I seesaw} mechanism~\cite{Minkowski:1977sc, mohapatra:1979ia, Yanagida:1979as, seesaw:1979, Schechter:1980gr}. Here the intermediate heavy particles are  clearly fermion singlets, identified as the RH Majorana neutrinos $N_{R\alpha}$.

\item [(ii)] The product of $L_\ell$ and $L_{\ell'}$ forms a scalar triplet: $(L_\ell^{\sf T}\sigma_a L_{\ell'})(\Phi^{\sf T}\sigma_a\Phi)/\Lambda$, where $\sigma_a$'s are the usual Pauli matrices. This is known as the {\em Type-II seesaw} mechanism~\cite{Schechter:1980gr, Magg:1980ut, Cheng:1980qt, lazarides:1980nt, mohapatra:1981yp}. Here the intermediate heavy particle is a scalar triplet $\mathbf{\Delta}=(\Delta^{++},\Delta^+,\Delta^0)$.

\item [(iii)] The product of $L_\ell$ and $\Phi$ forms a fermion triplet: $(L_\ell^{\sf T}{\sigma}^a\Phi)(L_{\ell'}^{\sf T}{\sigma}^a\Phi)/\Lambda$.  This is  known as the {\em Type-III seesaw} mechanism~\cite{foot:1988aq}. Here the intermediate heavy particle is a fermion triplet $\mathbf{\Sigma}=(\Sigma^{+},\Sigma^0,\Sigma^-)$.   
\end{enumerate}
One could also construct a hybrid seesaw model using more than one of these different types of seesaw mechanism. Non-minimal variations of the seesaw mechanism with higher multiplets have also been constructed; see e.g.~\cite{Picek:2009is, Kumericki:2012bh}. 

The origin of the bare Majorana mass terms responsible for the {\em explicit} $B-L$ violation can be understood from natural implementations of the seesaw mechanism in ultra-violet (UV) complete theories, e.g. in the Left-Right Symmetric Model (LRSM)~\cite{Mohapatra:1974hk, Mohapatra:1974gc, Senjanovic:1975rk} and in $SU(5)$~\cite{Bajc:2006ia}, $SO(10)$~\cite{Witten:1979nr, Babu:1992ia} and $E_6$~\cite{mohapatra:1986aw, Nandi:1985uh, mohapatra:1986bd} Grand Unified Theories (GUTs); for a review, see e.g.~\cite{ross}.

For the conventional seesaw scenarios mentioned above, the light neutrino masses are inversely proportional to a large lepton-number breaking scale (hence the name `seesaw'). There exists an important variation, where the light neutrino masses are {\em directly} proportional to a small lepton-number breaking scale. This is known as the {\em inverse seesaw} mechanism~\cite{mohapatra:1986bd} (see also~\cite{mohapatra:1986aw, Nandi:1985uh}) and can be regarded as an extension of the Type-I seesaw, since we introduce two sets of SM singlet fermions, instead of one. Experimentally, the main distinguishing feature of this scenario is the {\em pseudo-Dirac} nature of the heavy SM-singlet fermions, in contrast with the purely {\em Majorana} nature of the singlet fermions in the Type-I seesaw scenario.    

Another interesting class of neutrino mass models uses the radiative mechanism which, unlike the tree-level seesaw models, can go beyond the effective dimension-5 operator and generate tiny neutrino masses at an arbitrary loop-level. Some of the simplest and predictive radiative seesaw models are given in~\cite{zee:1980ai, zee:1985id, babu:1988ki, Pilaftsis:1991ug, ma:1998dn, Ma:2006km, FileviezPerez:2009ud, FileviezPerez:2010ch, 
Dev:2012sg, Law:2013dya,  Babu:2013pma}.  
In supersymmetric theories, there exists yet another possibility of generating non-zero neutrino masses without the need of a seesaw mechanism. The minimal way to incorporate $(B-L)$-breaking with the MSSM particle content is by bilinear $R$-parity violation~\cite{Hall:1983id, Joshipura:1994ib, Smirnov:1995ey, Nowakowski:1995dx, Hempfling:1995wj, Nilles:1996ij, Diaz:1997xc, Joshipura:1998fn, Kaplan:1999ds, Hirsch:2000ef}. For a review on various low-scale neutrino mass models, see e.g.~\cite{Boucenna:2014zba}. 

In a bottom-up approach, the scale of new physics responsible for $(B-L)$-breaking is {\em a priori} unknown. Since this review is about the collider tests of neutrino physics, we will only consider those scenarios with a $(B-L)$-breaking scale $\lesssim {\cal O}({\rm TeV})$ accessible to foreseeable collider experiments. The rest of the review is organized as follows: in Section~\ref{sec:2}, we discuss the collider prospects of sterile neutrinos appearing in the minimal Type-I seesaw and its singlet extensions within the SM gauge group. In Section~\ref{sec:3}, we discuss the collider prospects of minimal Type-II and Type-III seesaw models. In Section~\ref{sec:4}, we review some seesaw models with extended gauge sectors, namely, with an additional $U(1)$ in Section~\ref{sec:4.1} and with the Left-Right (L-R) Symmetric gauge group in Section~\ref{sec:4.2}.  In Section~\ref{sec:5}, we discuss the significance of the observation of LNV for leptogenesis. In Section~\ref{sec:6}, we briefly discuss the supersymmetric neutrino mass models. Finally, in Section~\ref{sec:7}, we present our conclusions and future prospects.

\section{Heavy Sterile Neutrinos at Colliders} \label{sec:2}
The simplest renormalizable extension of the SM for understanding the smallness of the LH neutrino masses is defined by the interaction Lagrangian 
\begin{align}
-{\cal L}_Y \ = \ h_{\ell \alpha} \bar{L}_\ell \widetilde{\Phi} N_{R\alpha} + {\rm H.c.} \;, 
\label{lagY}
\end{align}
where $\widetilde \Phi=i\sigma_2\Phi^*$ and $N_{R\alpha}$ are SM singlet neutral fermions, also known as the sterile neutrinos, since they cannot directly participate in the SM charged-current (CC) and neutral-current (NC) interactions in the absence of any mixing with the active neutrino sector. In \eqref{lagY}, 
$\alpha=1,2,\cdots,{\cal N}$ is the sterile neutrino flavor index and $h_{\ell \alpha}$ are the dimensionless complex Yukawa couplings.  
From the structure of~\eqref{lagY}, we see that the fermions $N_\alpha$ must necessarily be right-chiral; hence, they are also known as RH neutrinos. This distinction will be naturally justified in the LRSM (see Section~\ref{sec:4.2}). 
Assuming that \eqref{lagY} is the {\em only} source of neutrino masses and oscillations, we need at least two or three RH neutrinos, depending on whether the lightest active neutrino is massless or not.\footnote{The current upper limits on the absolute active neutrino mass scale from the kinematics of tritium $\beta$-decay~\cite{Kraus:2004zw, Aseev:2011dq}, as well as the cosmological bounds on the sum of neutrino masses~\cite{Planck:2015xua} still allow for three non-zero active neutrinos.}

In the Higgs phase after EWSB, the term \eqref{lagY} generates a Dirac mass $M_D=hv$. Since the RH neutrinos carry no SM gauge charges, one can also write a Majorana mass term 
\begin{align}
-{\cal L}_M \ = \ \frac{1}{2}(M_N)_{\alpha \beta} \bar{N}_{R\alpha}^C N_{R\beta} +{\rm H.c.} \; ,
\label{lagM}
\end{align}
while preserving gauge invariance. The term \eqref{lagM} implies that the hypercharge of $N_{R\alpha}$ is zero, and therefore, from \eqref{lagY}, we deduce that the hypercharges of the lepton and Higgs doublets are the same. Thus, the requirement of cancellation of gauge chiral anomalies implies charge quantization, provided that the neutrino mass eigenstates are Majorana fields~\cite{Babu:1989tq, Babu:1989ex, Foot:1990uf, Nowakowski:1992ff}.\footnote{This is true regardless of the number of generations, in contrast with the SM (without RH neutrinos) where anomaly cancellation implies charge quantization {\em only} in the one generation case~\cite{Deshpande:1979nr, Geng:1988pr}.}

The terms \eqref{lagY} and \eqref{lagM} together lead to the following neutrino mass matrix in the flavor basis $\{\nu^C_{L\ell},N_{R\alpha}\}$:
\begin{align}
\label{eq:seesaw}
	{\cal M}_\nu \ = \ 		\begin{pmatrix}
			\mathbf{0}   & M_D \\
				M_D^{\sf T} & M_N
		\end{pmatrix} \; .
\end{align}
For $\|M_DM_N^{-1}\|\ll 1$ (with $\|M\|\equiv \sqrt{{\rm Tr}(M^\dag M)}$ being the norm of matrix $M$), the light neutrino masses and mixing are given by the diagonalization of the effective mass matrix 
\begin{align}
M_\nu \ \simeq \ - M_D M_N^{-1} M_D^{\sf T} \; ,
\label{Mnu}
\end{align}
and the left-right neutrino mixing parameter is given by $V_{\ell N_\alpha}\sim M_DM_N^{-1}$. This is the Type-I seesaw mechanism~\cite{Minkowski:1977sc, mohapatra:1979ia, Yanagida:1979as, seesaw:1979, Schechter:1980gr}, as mentioned in Section~\ref{sec:1}. 

From the above discussion, it is clear that there are two key aspects of the Type-I seesaw mechanism that can be probed experimentally: the Majorana mass $M_N$ of the mostly sterile neutrinos and their mixing $V_{\ell N}$ with the active neutrinos. The Majorana nature of both light and heavy neutrinos can in principle be tested via the classic LNV process of neutrinoless double beta decay ($0\nu\beta\beta$); for reviews, see e.g.~\cite{Rodejohann:2011mu, Bilenky:2014uka}. However, an observation of $0\nu\beta\beta$ does not necessarily provide us with information on the mixing $V_{\ell N}$, whose effects on the $0\nu\beta\beta$ amplitude may be sub-dominant, compared to purely left- or right-chiral contributions. On the other hand, if the mixing effects are non-negligible, they could be inferred from non-unitarity of the light neutrino mixing matrix~\cite{Antusch:2006vwa, abada:2007ux, Antusch:2014woa}, as well as in low-energy observables for lepton flavor violation (LFV), lepton non-universality and electroweak precision tests~\cite{delAguila:2008pw, Akhmedov:2013hec, Basso:2013jka, Blas:2013ana, Antusch:2014woa, Antusch:2015mia}. However, these low-energy observables by themselves do not prove the Majorana nature of heavy neutrinos, since models with pseudo-Dirac heavy neutrinos can also yield large non-unitarity and LFV effects~\cite{Malinsky:2009gw, malinsky:2009df, Dev:2009aw, Forero:2011pc, Awasthi:2013ff}. As we will discuss in this section, the collider experiments at the energy frontier provide a {\em simultaneous} probe of both the aspects of seesaw, if the heavy neutrinos are kinematically accessible. This is complementary to the low-energy searches of LNV and LFV at the intensity frontier. 

In a bottom-up approach, the RH neutrinos can just be introduced `by hand' as the only new particles beyond the SM, as e.g.~in the {\em Neutrino SM} ($\nu$SM)~\cite{Asaka:2005an, Asaka:2005pn}. In this case, the RH neutrino masses in \eqref{lagM} are largely unconstrained, even with the requirement of satisfying the neutrino oscillation data. For instance, if the Yukawa couplings $|h_{\ell \alpha}| \sim {\cal O}(1)$, one expects from \eqref{Mnu} that the Majorana mass scale $M_N\sim {\cal O}(10^{15}~{\rm GeV})$, which is close to the upper limit due to partial-wave unitarity: $\Lambda\lesssim 4\pi v^2/\sqrt 3 M_\nu$~\cite{Maltoni:2000iq}.  For any smaller value of $M_N$, one could find the associated values of Yukawa couplings, as required to fit the light neutrino data using the seesaw formula \eqref{Mnu}. In fact, there exist seesaw models with $M_N$ spanning over a wide range of scales, i.e. from eV to GUT scale (for a review, see~\cite{Drewes:2015jna}). However, from the experimental point of view, only the scenarios with $M_N\lesssim {\cal O}({\rm TeV})$ offer the possibility of being tested in foreseeable future. Even from a theoretical point of view, the naturalness requirement that electroweak corrections to the bilinear Higgs mass operator $\mu^2\Phi^\dag \Phi$ should not exceed $\delta \mu \sim {\cal O}(1~{\rm TeV})$ imposes an {\em upper} bound of $M_N\lesssim 4\times 10^7$ GeV on the lightest RH neutrino mass~\cite{Vissani:1997ys, casas:2004gh,  Farina:2013mla, Clarke:2015gwa}. Supergravity models of inflation impose an additional {\em upper} bound on the reheating temperature $T_R\lesssim 10^6-10^9$ GeV (and hence, on the RH neutrino masses), as required to avoid overproduction of gravitinos whose late decays may otherwise spoil the success of Big Bang Nucleosynthesis (BBN)~\cite{khlopov:1984pf, ellis:1984eq, Ellis:1984er, Kawasaki:1994af, Cyburt:2002uv, kawasaki:2004qu, Kawasaki:2008qe}. Another reason for considering the low-scale seesaw mechanism is that it provides a unique  opportunity to test the link between the neutrino mass mechanism and the observed matter-antimatter asymmetry in our Universe via the mechanism of leptogenesis (for a review, see e.g.~\cite{Davidson:2008bu}). Motivated by all these theoretical and experimental considerations, we will focus on the low-scale seesaw scenarios. 

We should also mention here that there has been a lot of recent interest in GeV-scale seesaw models, especially within the $\nu$SM framework~\cite{Asaka:2005an, Asaka:2005pn}, which can explain not only neutrino masses, but also matter-antimatter asymmetry and Dark Matter (DM), in a minimal setup (for a review, see e.g.~\cite{Drewes:2013gca}). This scenario relies on the possibility that two sterile neutrinos with masses typically in the range of 1-10 GeV, well below the critical temperature of the electroweak phase transition, lead to an enhanced lepton asymmetry via coherent $C\!P$-asymmetric oscillations~\cite{Akhmedov:1998qx, Asaka:2005pn, Shaposhnikov:2008pf, Canetti:2012kh, Shuve:2014zua}, which is of ${\cal O}(h^6)$ but could be sufficient to explain the observed matter-antimatter asymmetry. This requires a very specific choice of model parameters, but might be accessible to current and future experiments at the intensity and cosmic frontiers.  In particular, the lightest RH neutrino could play the role of a DM, if its mass is in the keV range (for a review, see e.g.~\cite{Merle:2013gea}). This has gained further attention in light of the recent observations of an unidentified $X$-ray line at energy of 3.5 keV~\cite{Bulbul:2014sua, Boyarsky:2014jta, Boyarsky:2014ska}, which could be explained as due to the radiative decay of a sterile neutrino DM with mass of 7 keV.

In an ultra-violet complete theory, such as in the LRSM~\cite{Mohapatra:1974hk, Mohapatra:1974gc, Senjanovic:1975rk} and in $SO(10)$ GUT~\cite{Fritzsch:1974nn}, the RH neutrinos constitute an integral part of the particle spectrum, as required by anomaly cancellation, and the RH neutrino mass scale is no longer an adhoc parameter, but intimately related to the $U(1)_{B-L}$-breaking scale. In the $SO(10)$-type theories, quark-lepton unification implies that the Yukawa couplings appearing in \eqref{lagY} are of similar order of magnitude as the up-quark Yukawa couplings, which means that the seesaw scale $M_N\sim 10^{14}$ GeV. Thus, even though the GUT seesaw models are quite elegant and predictive~\cite{Babu:1992ia}, testing their ultra-heavy particle spectrum experimentally, going beyond the `grand' desert, is a formidable task. On the other hand, simpler gauge-extended seesaw models, such as the L-R seesaw, could easily accommodate an experimentally-accessible RH neutrino mass scale. We will discuss some of these scenarios in Section~\ref{sec:4}.\footnote{An alternative low-scale LNV can be realized in models with {\em spontaneous} global $B-L$ violation, which implies the existence of a massless Goldstone boson called the Majoron~\cite{Chikashige:1980qk, chikashige:1981ui, Schechter:1981cv}  
with potentially interesting phenomenological consequences for $0\nu\beta\beta$~\cite{berezhiani:1992cd, Burgess:1992dt, Burgess:1993xh, Bamert:1994hb, Hirsch:1995in}, leptogenesis~\cite{Pilaftsis:2008qt}, DM~\cite{gelmini:1984pe, berezinsky:1993fm,  Queiroz:2014yna} and inflation~\cite{Boucenna:2014uma}. In the context of supersymmetry, both $B-L$ and $R$-parity can be spontaneously broken by the VEV of the RH sneutrino field, leading to a minimal model of $R$-parity violation~\cite{Barger:2008wn}. }   

\subsection{Low-Scale Singlet Seesaw Models}\label{sec:2.1}

In the traditional ``vanilla" seesaw mechanism~\cite{Minkowski:1977sc, mohapatra:1979ia, Yanagida:1979as, seesaw:1979, Schechter:1980gr}, the left-right neutrino mixing is given by  
\begin{align}
V_{\ell N} \ \simeq \ \sqrt{\frac{M_\nu}{M_N}} \ \lesssim \ 10^{-6} \sqrt{\frac{100~{\rm GeV}}{M_N}}\; ,
\label{canon}
\end{align}
due to the smallness of the light neutrino mass $M_\nu \lesssim 0.1~{\rm eV}$~\cite{Planck:2015xua}. Thus for a low seesaw scale in the sub-TeV to TeV range, the experimental effects of the light-heavy neutrino mixing are expected to be too small, unless the RH neutrinos have additional interactions, e.g. when they are charged under $U(1)_{B-L}$. However, there exists a class of minimal SM plus low-scale Type-I seesaw scenarios~\cite{Pilaftsis:1991ug, Buchmuller:1991ce, Gluza:2002vs, Pilaftsis:2004xx, Kersten:2007vk,  Xing:2009in, Gavela:2009cd, He:2009ua,  Adhikari:2010yt, Ibarra:2010xw, Deppisch:2010fr, Ibarra:2011xn, Mitra:2011qr}, where $V_{\ell N}$ can be sizable while still satisfying the light neutrino data. This is made possible by assigning specific textures to the Dirac and Majorana mass matrices in the seesaw formula \eqref{Mnu}. The stability of these textures can in principle be guaranteed by enforcing some symmetries in the lepton sector~\cite{Pilaftsis:2004xx, Shaposhnikov:2006nn, Kersten:2007vk, Deppisch:2010fr, Dev:2013oxa}. We will generically assume this to be the case for our subsequent discussion on the collider signatures of low-scale minimal seesaw, without referring to any particular texture or model-building aspects.  Also, unless otherwise specified, we will use a model-independent phenomenological approach, parametrized by a single heavy neutrino mass scale $M_N$ and a single flavor light-heavy neutrino mixing $V_{\ell N}$, assuming that the mixing effects in other flavors $\ell'\neq \ell$ are sub-dominant. Although this assumption may not be strictly valid for a realistic seesaw model satisfying the observed neutrino oscillation data, it enables us to derive generic bounds on the mixing parameter, which could be translated or scaled appropriately in the context of particular neutrino mass models (see e.g.~\cite{Drewes:2015iva}). 

Another natural realization of a low-scale seesaw scenario with large light-heavy neutrino mixing is the inverse seesaw model~\cite{mohapatra:1986bd}, where one introduces two sets of SM singlet fermions $\{N_{R\alpha}, S_{L\rho}\}$ with opposite lepton numbers, i.e. $L(N_R)=+1=-L(S_L)$. In this case, the neutrino Yukawa  sector of  the Lagrangian is in general given by 
\begin{align}
-{\cal L}_Y \ = \ h_{l\alpha}\bar L_\ell \widetilde{\Phi} N_{R\alpha}
+ (M_{S})_{\rho\alpha} \bar{S}_{L\rho}  N_{R\alpha} 
+ \frac{1}{2}\left[(\mu_R)_{\alpha\beta} \bar{N}^C_{R\alpha} 
N_{R\beta} + (\mu_S)_{\rho\lambda}\bar{S}_{L\rho}
 S^C_{L\lambda} \right] + {\rm H.c.}\; ,
\label{lag_inv} 
\end{align}
where $M_S$ is a Dirac mass term and $\mu_{R,S}$ are Majorana mass terms. After EWSB, the Lagrangian \eqref{lag_inv} gives rise to the following neutrino mass matrix 
in the flavor basis $\{(\nu_{L\ell})^C,N_{R\alpha},(S_{L\rho})^C \}$: 
\begin{align}
{\cal M}_\nu\ =\ \left(\begin{array}{ccc}
{\bf 0} & M_D & {\bf 0}\\
M_D^{\sf T} & \mu_R &  M_S^{\sf T}\\
{\bf 0} & M_S & \mu_S 
\end{array}\right) \ \equiv \ \left(\begin{array}{cc} \mathbf{0} & {\cal M}_D \\ {\cal M}_D^{\sf T} & {\cal M}_N \end{array}\right)\; ,
\label{eq:inverse1}
\end{align}
which has a form similar to the Type-I seesaw matrix \eqref{eq:seesaw}, with ${\cal M}_D = (M_D, \mathbf{0})$ and ${\cal M}_N = \left(\begin{array}{cc} \mu_R & M_S^{\sf T} \\ M_S & \mu_S \end{array}\right)$. 
Here we have not considered the dimension-4 lepton-number breaking term $\bar L \widetilde{\Phi} S_L^C$ which appears, for instance, in linear seesaw models~\cite{wyler:1983dd, akhmedov:1995ip, akhmedov:1995vm, malinsky:2005bi}, since the mass matrix in presence of this term can always be rotated to the form given by (\ref{eq:inverse1})~\cite{Ma:2009du}. Also observe that  the  inverse seesaw model  discussed originally
in~\cite{mohapatra:1986bd} set the RH neutrino Majorana mass 
$\mu_R = {\bf 0}$ in ~(\ref{eq:inverse1}). At the tree-level, the light neutrino mass is directly proportional to the Majorana mass term $\mu_S$ for $\|\mu_S\| \ll \|M_S\|$:
\begin{align}
M_\nu \ = \ M_D M_S^{-1} \mu_S M_S^{-1^{\sf T}} M_D^{\sf T} + {\cal O}(\mu_S^3),
\label{Mnu_inv}
\end{align}
 whereas at the one-loop level, there is an additional contribution proportional to $\mu_R$~\cite{Dev:2012sg, Dev:2012bd}, arising from standard electroweak radiative corrections~\cite{Pilaftsis:1991ug}. The smallness
of $\mu_{R,S}$ is  `technically natural' in
the 't Hooft sense~\cite{thooft:1979}, i.e. in the limit of $\mu_{R,S}\to {\bf 0}$,
lepton   number  symmetry   is  restored   and  the   light  neutrinos
$\nu_{L\ell}$  are massless  to all  orders in  perturbation theory, as in the SM. 

The freedom provided by the small LNV parameter $\mu_S$ in \eqref{Mnu_inv} is the key feature of the inverse seesaw mechanism, allowing us to fit the light neutrino data for {\em any} value of light-heavy neutrino mixing, without introducing any fine-tuning or cancellations in the light neutrino mass matrix \eqref{Mnu_inv}~\cite{gonzalezgarcia:1988rw, bernabeu:1987gr}.  In essence, the magnitude of the neutrino mass becomes decoupled from the heavy neutrino mass, thus allowing for a large mixing 
\begin{align}
\label{eq:thetainvseesaw}
		V_{\ell N} \simeq \ \sqrt{\frac{M_\nu}{\mu_S}} \ \approx \ 10^{-2}\sqrt{\frac{1~\text{keV}}{\mu_S}} \; .
\end{align}
The heavy neutrinos $N_R$ and $S_L$ have opposite $C\!P$ parities and form a quasi-Dirac state with relative mass splitting of the order $\kappa = \mu_S/M_S$. All LNV processes are usually suppressed by this small mass splitting. For instance, in the one-generation case, the light neutrino mass in \eqref{Mnu_inv} can be conveniently expressed as $M_\nu \simeq \kappa V_{\ell N} M_D$, in contrast with $V_{\ell N}M_D$ in the Type-I seesaw case [cf.~\eqref{Mnu}]. It should be noted here that the approximately $L$-conserving models with quasi-degenerate heavy Majorana neutrinos could provide a natural framework~\cite{Asaka:2008bj, GonzalezGarcia:2009qd, Blanchet:2009kk, Blanchet:2010kw} for realizing the mechanism of resonant leptogenesis\cite{Pilaftsis:1997dr, Pilaftsis:1997jf, pilaftsis:2003gt}, where the leptonic $C\!P$ asymmetry is resonantly enhanced when the mass splitting $\Delta M_N$ is of the same order as the decay width $\Gamma_N$.
\begin{figure}[t]
\centering
\includegraphics[clip,width=0.45\textwidth]{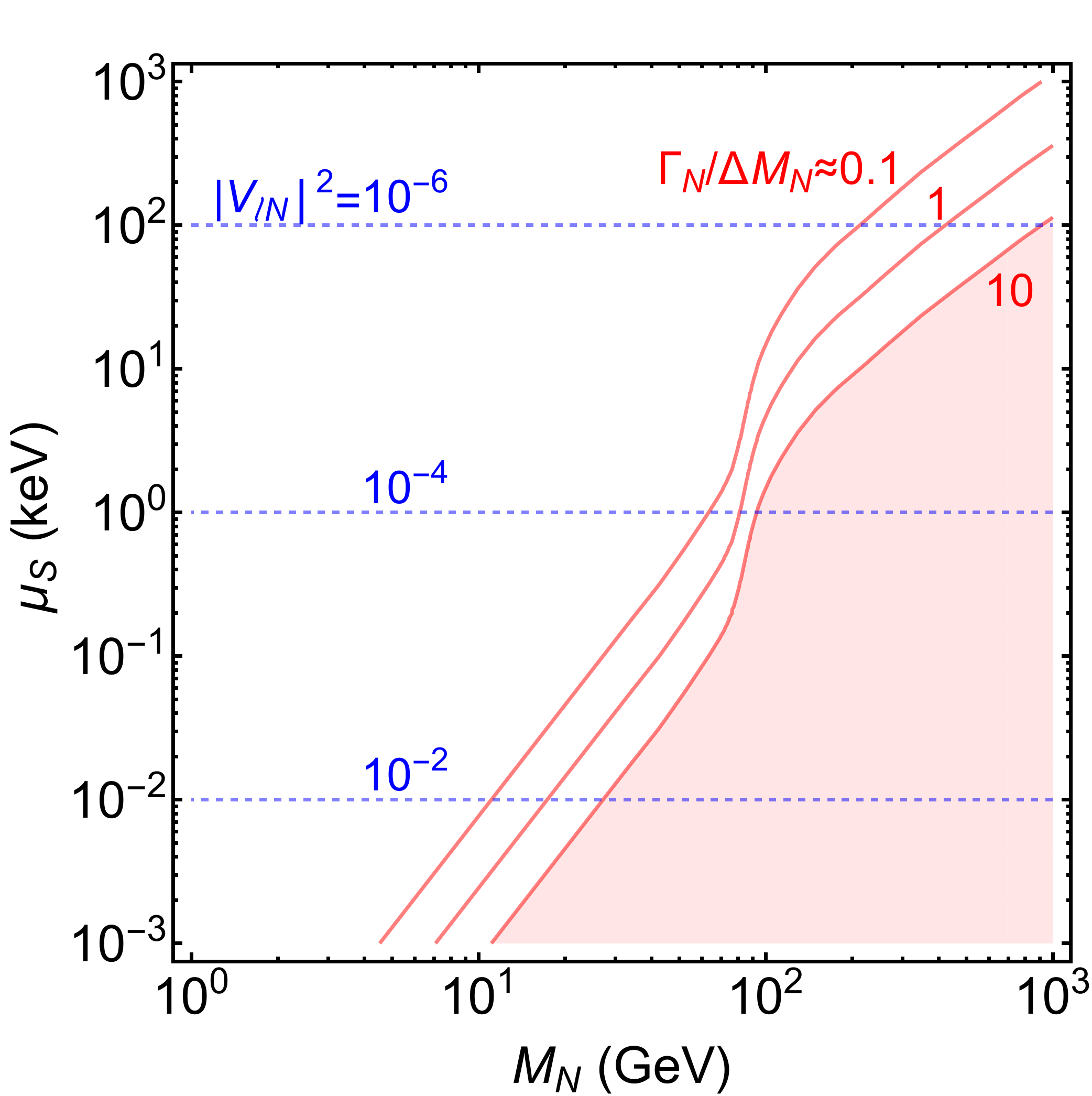}
\caption{Contours of the ratio of the average decay width of the heavy neutrinos to their mass splitting in the inverse seesaw model. The LNV signal will be unobservable in the shaded region with $\Gamma_N/\Delta M_N\gtrsim 10$. }
\label{fig:width}
\end{figure}

As for the LNV signature at colliders, in a natural seesaw scenario with approximate lepton number conservation, the LNV amplitude for the on-shell production of heavy neutrinos at average four-momentum squared $\bar{s}=(M_{N_1}^2+M_{N_2}^2)/2$ can be written as 
\begin{align}
{\cal A}_{\rm LNV}(\bar{s}) \ = \ -V_{\ell N}^2\frac{2\Delta M_N}{\Delta M_N^2+\Gamma_N^2}+{\cal O}\left(\frac{\Delta M_N}{M_N}\right) \; ,
\label{ampLNV}
\end{align}
for $\Delta M_N\lesssim \Gamma_N$, i.e. for small mass difference $\Delta M_N=|M_{N_1}-M_{N_2}|$ between the heavy neutrinos compared to their average decay width $\Gamma_N\equiv (\Gamma_{N_1}+\Gamma_{N_2})/2$. Thus, the LNV amplitude in \eqref{ampLNV} will be suppressed by the small mass splitting, except for the case $\Delta M_N\simeq \Gamma_N$ when it can be resonantly enhanced~\cite{Pilaftsis:1997dr, Bray:2007ru}. This suppression for the inverse seesaw case is illustrated in Figure~\ref{fig:width} where we show the contours of $\Gamma_N/\Delta M_N$ for different values of the inverse seesaw mass parameters $M_N$ and $\mu_S$. As we can see from the plot, to observe LNV with $M_N\gtrsim {\cal O}(100)$ GeV, one needs a large $\mu_S$, and therefore, small $|V_{\ell N}|^2$ [cf.~\eqref{eq:thetainvseesaw}], which will suppress the LNV signal. Its implications for collider searches of these scenarios will be discussed in Section~\ref{sec:direct}.

\begin{figure}[t]
\centering
\includegraphics[width=7.7cm]{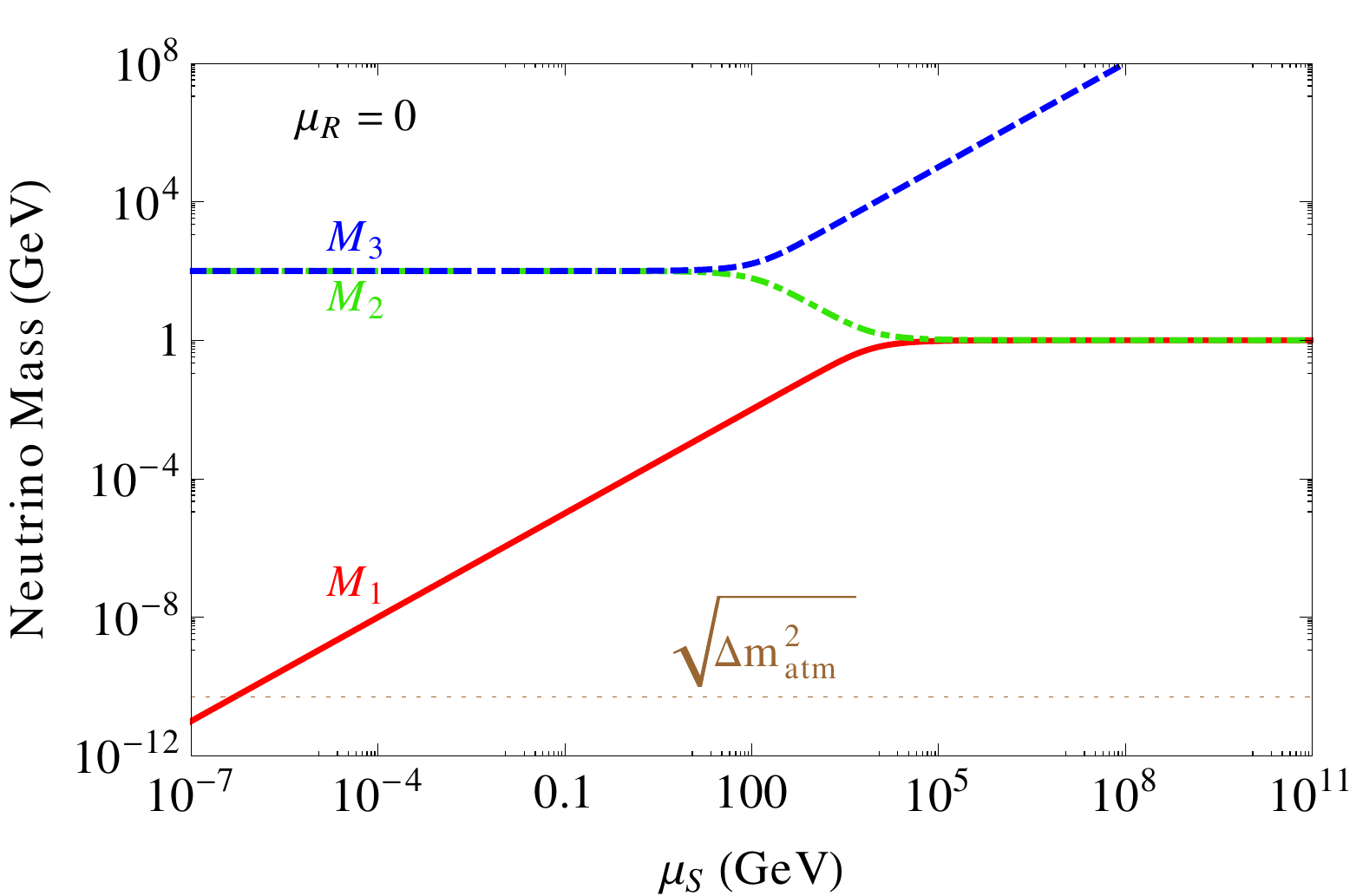}
\includegraphics[width=7.7cm]{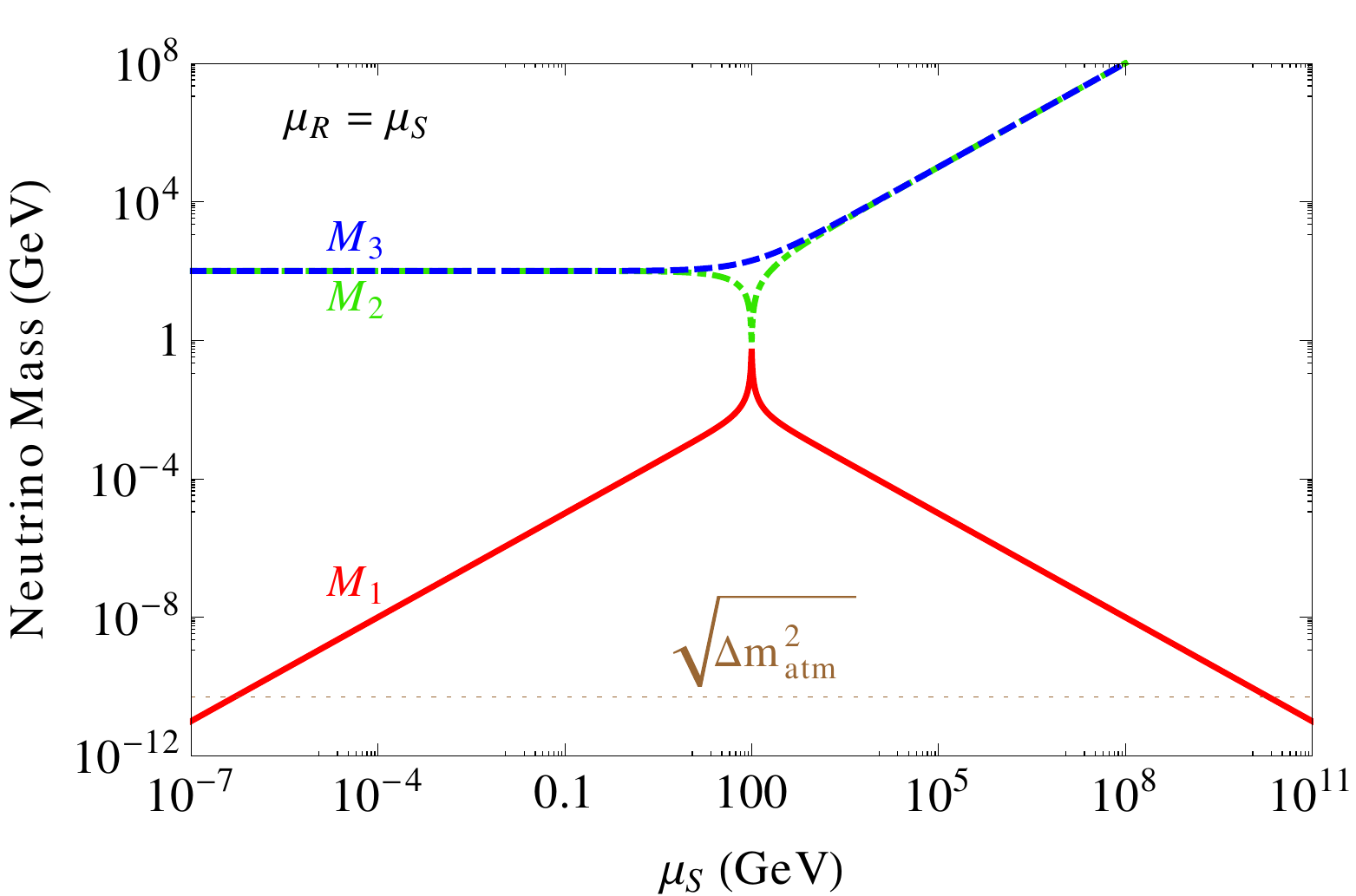}
\caption{Absolute eigenvalues of the neutrino mass matrix \eqref{eq:inverse1} as a function of the LNV mass term $\mu_S$ in the one-generation case with $\mu_R=0$ (left panel) and $\mu_R=\mu_S$ (right panel). Here we have chosen $M_D = 1$~GeV and $M_S = 100$~GeV for illustration. The horizontal dotted line shows the light neutrino mass scale required to satisfy the atmospheric neutrino data.}
\label{fig:spectrum} 
\end{figure}
By virtue of the last equivalence in \eqref{eq:inverse1}, the inverse seesaw mechanism can be regarded as a variation of the Type-I seesaw mechanism. Therefore, it is instructive to study the transition between the neutrino mass formulas given by \eqref{Mnu} and \eqref{Mnu_inv}. For illustration purposes, let us take a simplified version of \eqref{eq:inverse1} for a single  
generation case. The neutrino mass spectrum for this scenario is shown in Figure~\ref{fig:spectrum} as a function of the LNV parameter $\mu_S$. As an example, we have chosen the Dirac mass term $M_D=1$ GeV and the heavy neutrino mass term $M_S=100$ GeV such that $V_{\ell N}\simeq M_D/M_S=10^{-2}$ is consistent with the current upper limit set by the electroweak precision data (EWPD)~\cite{Blas:2013ana}. First we consider the original inverse seesaw model with $\mu_R=0$ in \eqref{eq:inverse1}. As shown in Figure~\ref{fig:spectrum} (left panel), successful light neutrino mass generation occurs in this case 
only for $\mu \sim 1$ keV [cf.~\eqref{eq:thetainvseesaw}]. As evident from \eqref{Mnu_inv}, for 
$\mu_S\ll M_S$, the lightest neutrino mass is proportional to $\mu_S$, whereas the two heavy neutrinos form a quasi-degenerate Dirac pair with masses $M_S\pm \mu_S$. For $\mu_S\gg M_S$, the heaviest neutrino mass becomes equal to $\mu_S$ and the lighter ones form a quasi-degenerate Dirac pair with mass of order $M_D$. For $\mu_R\neq 0$, the situation remains unchanged for the general case $\mu_{R,S}\ll M_S$, as shown in Figure~\ref{fig:spectrum} (right panel). However, for $\mu_{R,S}\gg M_S$, we recover the Type-I seesaw with the lightest neutrino mass given by $-M_D^2/\mu_R$ [cf.~\eqref{Mnu}], whereas the heavier ones form a Majorana pair with masses equal to $\mu_{R,S}$.

\subsection{Experimental Searches}\label{sec:2.2}
Various laboratory searches have put stringent constraints on sterile neutrino mixing with active ones in a wide mass range of $M_N$ from eV - TeV. For $M_N$ values well below 1 MeV, as e.g. in eV-seesaw models~\cite{deGouvea:2005er}, the sterile neutrinos can be probed by neutrino-oscillation experiments. Assuming all sterile neutrinos to be of the same order of magnitude, current data rule out $1~{\rm neV} \lesssim M_N \lesssim 1~{\rm eV}$~\cite{Cirelli:2004cz, deGouvea:2005er, deGouvea:2009fp, Donini:2012tt}. For $10~{\rm eV}\lesssim M_N \lesssim 1~{\rm MeV}$, the mixing of sterile neutrinos with electron neutrino has been constrained by searches for $0\nu\beta\beta$ and precision measurements of $\beta$-decay energy spectra. For $1~{\rm MeV}\lesssim M_N \lesssim 1~{\rm GeV}$, the mixing with both electron and muon neutrinos have been constrained by peak searches in leptonic decays of pions and kaons. Sterile neutrino mixing with all neutrino flavors in the MeV-GeV mass range has also been searched for through their decay products in beam dump experiments. 
Upper limits on the active-sterile neutrino mixing elements have also been derived from cosmological bounds on sterile neutrino lifetimes as required for the success of BBN~\cite{Gorbunov:2007ak, Boyarsky:2009ix, Ruchayskiy:2012si}. 

\begin{figure}[t]
\centering
\includegraphics[width=15cm]{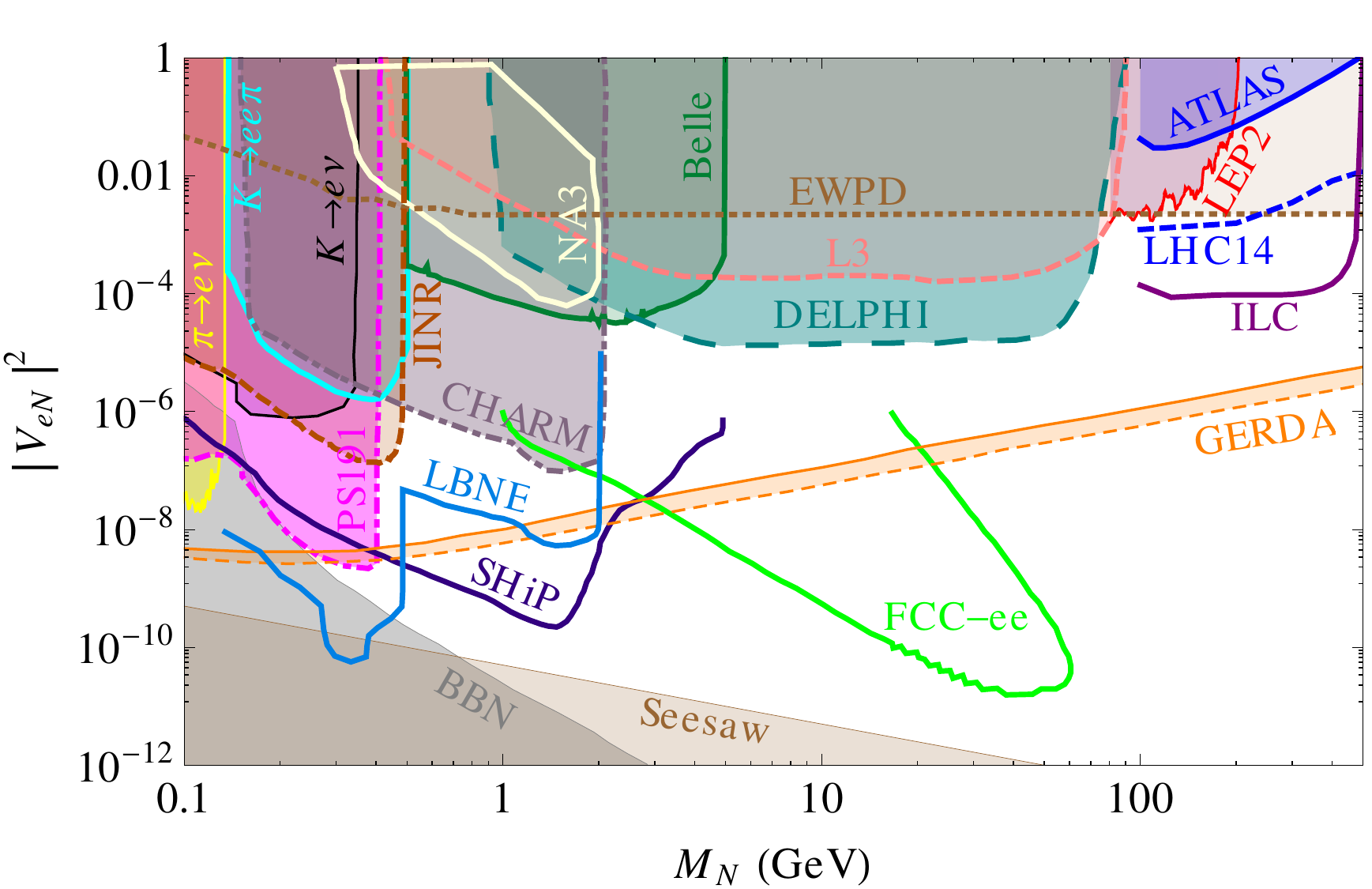}
\caption{Limits on the mixing between the electron neutrino and a single heavy neutrino in the mass range 100 MeV - 500 GeV. For details, see text. 
}
\label{fig:limVeN}
\end{figure}

\begin{figure}[t]
\centering
\includegraphics[width=15cm]{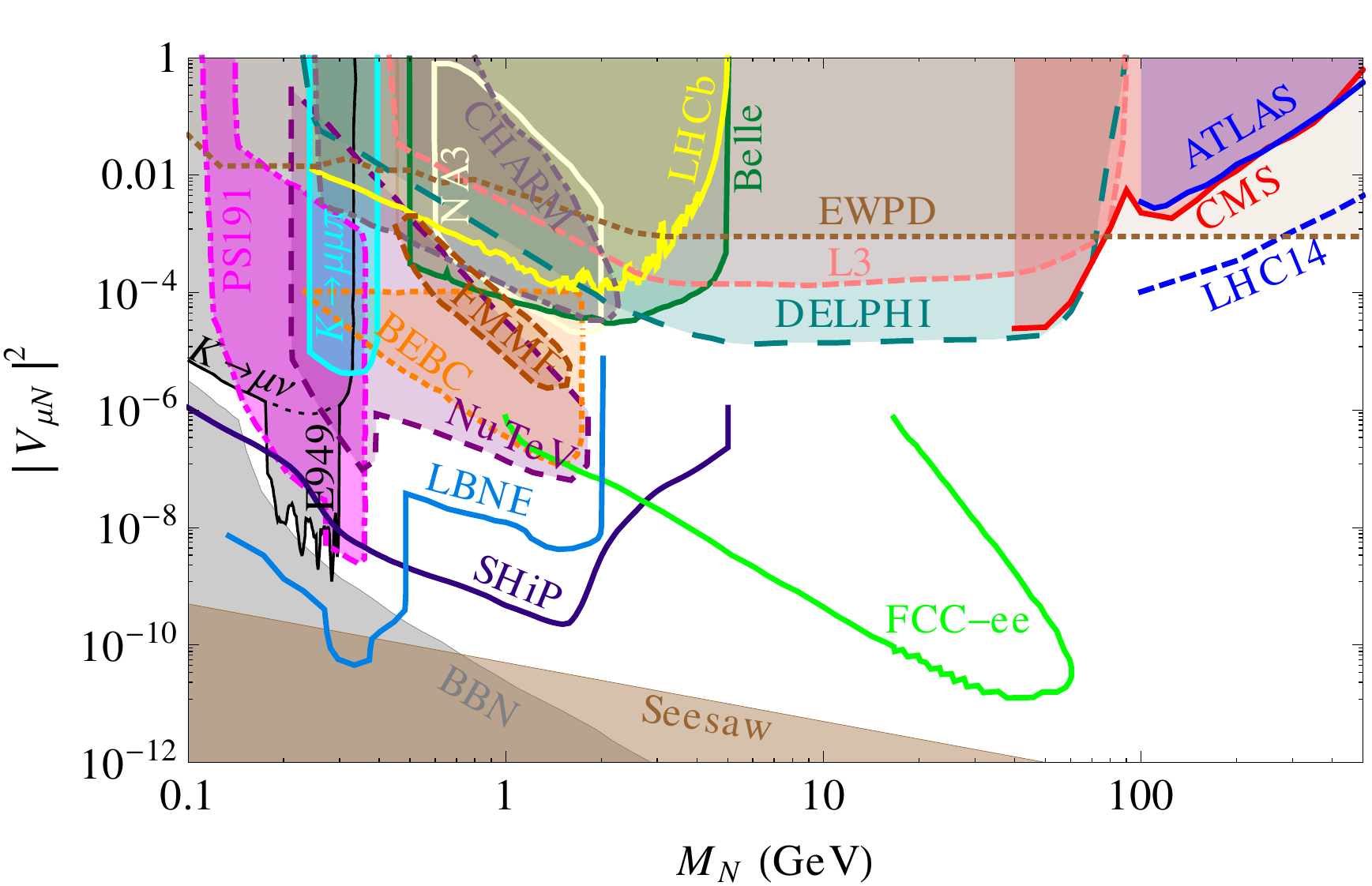}
\caption{Limits on the mixing between the muon neutrino and a single heavy neutrino in the mass range 100 MeV - 500 GeV. For details, see text.
}
\label{fig:limVmuN}
\end{figure}

\begin{figure}[t]
\centering
\includegraphics[width=15cm]{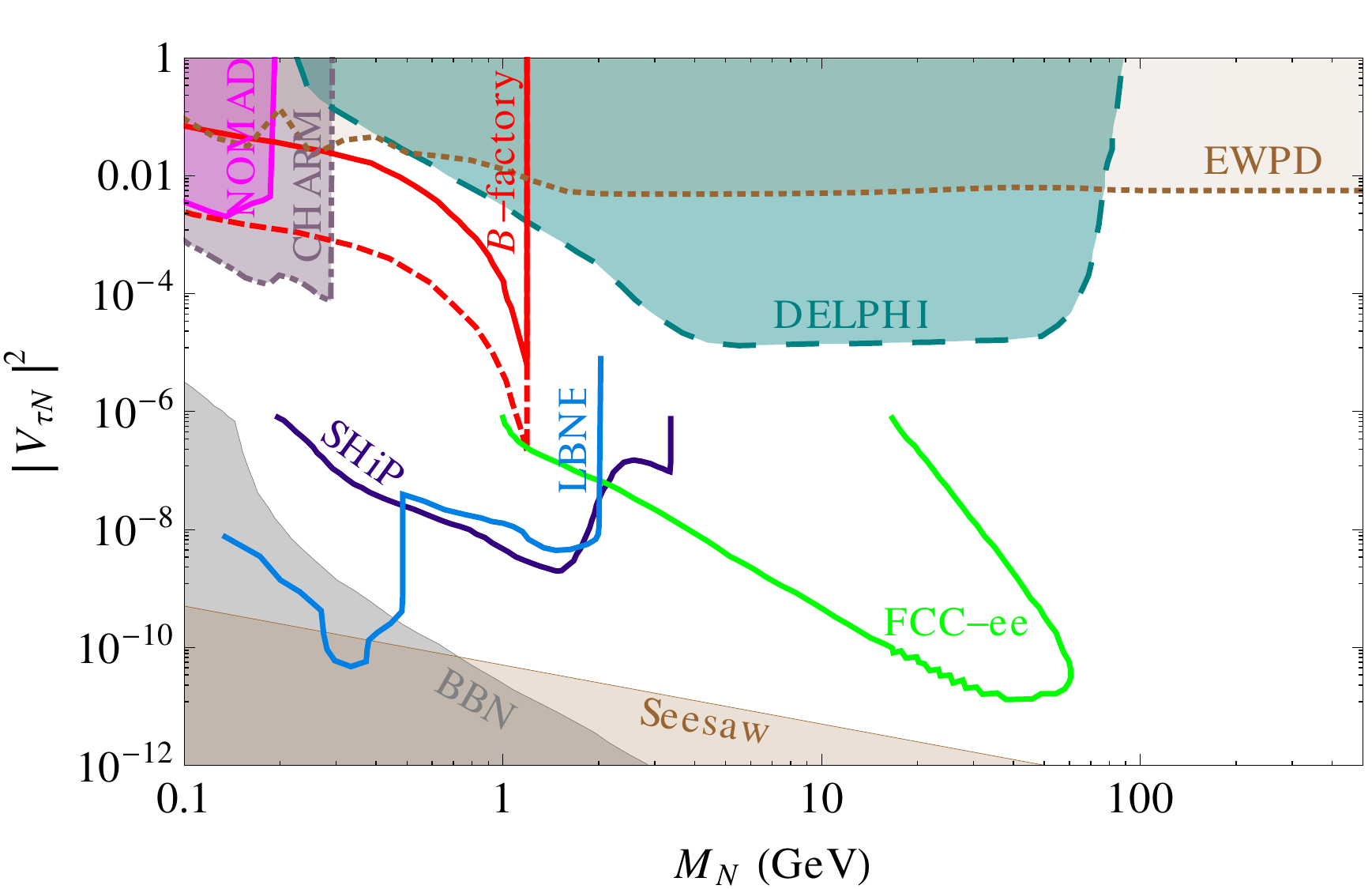}
\caption{Limits on the mixing between the tau neutrino and a single heavy neutrino in the mass range 100 MeV - 500 GeV. For details, see text. 
}
\label{fig:limVtauN}
\end{figure}

Here we summarize the current state-of-the-art sterile neutrino searches in the mass range $100~{\rm MeV}\leq M_N \leq 500~{\rm GeV}$, as relevant for collider experiments at the energy frontier and other planned experiments at the intensity frontier. Figures~\ref{fig:limVeN}-\ref{fig:limVtauN} show the current constraints and some future projections on sterile neutrino mixing with the electron, muon and tau neutrinos, respectively. In these plots, the (gray) contour labeled `BBN' corresponds to a heavy neutrino lifetime $> 1$ sec, which is disfavored by BBN constraints~\cite{Gorbunov:2007ak, Boyarsky:2009ix, Ruchayskiy:2012si}. The (brown) line labeled `Seesaw' shows the scale of mixing as expected in the canonical seesaw [cf.~\eqref{canon}]. We should remember that both these limits may get substantially modified in presence of more than two heavy neutrinos~\cite{Drewes:2015iva}. Other limits shown in Figures~\ref{fig:limVeN}-\ref{fig:limVtauN} are explained below. 


\subsubsection{Neutrinoless Double Beta Decay}\label{sec:0nbb}
The contributions of heavy Majorana neutrinos $N_{R\alpha}$ to $0\nu\beta\beta$ amplitude is described by the standard neutrino exchange diagram between two $\beta$-decaying neutrons, via a non-zero admixture of a $\nu_{Le}$ weak eigenstate parametrized by the mixing element $V_{eN_\alpha}$. 
The $0\nu\beta\beta$ half-life is given by 
\begin{align}
\frac{1}{T_{1/2}^{0\nu}} \ = \  {\cal A}\left|\frac{m_p}{\langle p^2 \rangle}\sum _{i=1}^3 U_{ei}^2 m_{i} \: + \: m_p\sum_{\alpha=1}^{{\cal N}} \frac{V_{eN_\alpha}^2 M_\alpha}{\langle p^2 \rangle+M_\alpha^2} \right|^2 \; ,
\label{half-life}
\end{align}
where ${\cal A}=G^{0\nu}g_A^4|{\cal M}_N^{0\nu}|^2$ and $\langle p^2 \rangle = m_p m_e|{\cal M}_N^{0\nu}/{\cal M}_\nu^{0\nu}|^2$. Here $m_p$ and $m_e$ are the proton and electron masses respectively, $G^{0\nu}$ is the phase-space factor, $g_A$ is the nucleon axial-vector coupling constant, $U_{ei}$ is the light neutrino mixing matrix, $m_i$ and $M_\alpha$ are respectively the light and heavy neutrino mass eigenvalues, and ${\cal M}_\nu^{0\nu}, {\cal M}_N^{0\nu}$ are the corresponding nuclear matrix elements  (NMEs). These NMEs are conventionally calculated for the limiting cases $m_i\ll p_F$ (light) and $M_\alpha \gg p_F$ (heavy), $p_F$ being the characteristic momentum transferred via the virtual neutrino, which is $\sim 200~{\rm MeV}$ corresponding to the mean nucleon momentum of Fermi motion in a nucleus. However, the interpolating formula~\eqref{half-life} allows us to calculate the $0\nu\beta\beta$ half-life for arbitrary heavy neutrino masses using the NMEs ${\cal M}_\nu^{0\nu}$ (light) and ${\cal M}_N^{0\nu}$ (heavy)~\cite{Kovalenko:2009td, Faessler:2014kka}. 

Using the combined 90\% C.L. limit on $0\nu\beta\beta$ half-life $T_{1/2}^{0\nu}(^{76}{\rm Ge})\geq 3\times 10^{25}~{\rm yr}$ from GERDA+Heidelberg-Moscow experiment~\cite{Agostini:2013mzu}, we derive from the second term in~\eqref{half-life} upper limits on $|V_{eN}|^2$ as a function of a generic heavy neutrino mass $M_N$. Our results are shown in Figure~\ref{fig:limVeN}, where the shaded (orange) region between the solid and dashed lines, labeled `GERDA', shows the uncertainty due to NMEs~\cite{Dev:2013vxa, Faessler:2014kka}.  Here we have used the recently re-evaluated phase-space factors~\cite{Kotila:2012zza} and the NMEs from a recent calculation within the quasi-particle random phase approximation (QRPA)~\cite{Simkovic:2013qiy, Faessler:2014kka}. Similar limits are obtained using the half-life limit $T_{1/2}^{0\nu}(^{136}{\rm Xe})\geq 2.6\times 10^{25}~{\rm yr}$ from KamLAND-Zen experiment~\cite{Gando:2012zm, TheKamLAND-Zen:2014lma} and the corresponding QRPA NMEs~\cite{Faessler:2014kka}. 

From Figure~\ref{fig:limVeN}, it seems that the $0\nu\beta\beta$ constraints are very severe, thus shadowing the future prospects of observing LNV in other processes involving the electron channel. However, one must keep in mind that the $0\nu\beta\beta$ limits may be significantly weakened in certain cases when a cancellation between different terms in \eqref{half-life} may happen~\cite{Pascoli:2013fiz}, e.g. due to the presence of Majorana $C\!P$ phases. In general, the Majorana nature of neutrinos does not guarantee an {\em observable} $0\nu\beta\beta$  rate in all models~\cite{Helo:2015fba}. Also, in the inverse seesaw scenario with pseudo-Dirac heavy neutrinos, the $0\nu\beta\beta$ limits are usually diluted by the small LNV term $\kappa=\mu_S/M_S$.  Therefore, it is still important to include the electron channel while performing an independent direct search for heavy neutrinos at colliders.

\subsubsection{Peak Searches in Meson Decays}\label{sec:peak}
Peak searches in weak decays of heavy leptons and mesons are powerful probes of heavy neutrino mixing with all lepton flavors. The most promising are the two-body decays of electrically charged mesons into leptons and neutrinos: $X^\pm \to \ell^\pm N$~\cite{Shrock:1980vy, Shrock:1980ct, Lello:2012gi}, whose branching ratio is proportional to the mixing $|V_{\ell N}|^2$. Thus, for a non-zero mixing and in the meson's rest frame, one expects the lepton spectrum to show a second monochromatic line at 
\begin{align}
E_{\ell} \ = \ \frac{M_X^2+m_\ell^2-M_N^2}{2M_X} \; ,
\label{peak}
\end{align} 
apart from the usual peak due to the active neutrino $\nu_{L\ell}$. For sterile neutrinos heavier than the charged lepton, the helicity suppression factor inherent in leptonic decay rate is weakened by a factor $M_N^2/m_\ell^2$~\cite{Shrock:1980ct} due to which the sensitivity on $|V_{\ell N}|^2$ {\em increases} with $M_N$ till the phase space becomes relevant. Peak searches have been performed in the channels $\pi \to eN$~\cite{Azuelos:1986eg, DeLeenerRosier:1991ic, Britton:1992pg, Britton:1992xv, PIENU:2011aa}, $\pi\to \mu N$~\cite{Abela:1981nf, Minehart:1981fv, Daum:1987bg, Bryman:1996xd, Assamagan:1998vy}, $K\to eN$~\cite{Yamazaki:1984sj} and $K\to \mu N$~\cite{Asano:1981he, Hayano:1982wu, Yamazaki:1984sj, Kusenko:2004qc, Artamonov:2014urb}. The current 90\% C.L. limits on $|V_{\ell N}|^2$ (for $\ell=e,\mu$) derived from these searches are shown in Figures~\ref{fig:limVeN} and \ref{fig:limVmuN}, labeled as `$X\to \ell \nu$' (with $X=\pi,K$ and $\ell=e,\mu$). The limit from $\pi\to \mu N$ is not shown here, since it is only applicable in the mass range 1 MeV $\leq M_N \leq $ 30 MeV.

The peak searches could in principle be extended to higher masses with heavier meson/baryon decays~\cite{Johnson:1997cj, Ramazanov:2008ph, Gninenko:2009yf}. 
For instance, the Belle experiment~\cite{Liventsev:2013zz} used the decay mode $B\to X\ell N$ followed by $N\to \ell \pi$ (with $\ell=e,\mu$) in a data sample of 772 million $B\bar{B}$ pairs coming from $\Upsilon(4s)$ resonance to place 90\% C.L. limits on $|V_{e N}|^2$ and $|V_{\mu N}|^2$ in the heavy neutrino mass range 500 MeV to 5 GeV, as shown in Figures~\ref{fig:limVeN} and \ref{fig:limVmuN}, labeled as `Belle'. 

Limits on the mixing parameter can also be set from the 3-body decay of muons, where a sterile neutrino contribution may distort the spectrum of Michel electrons~\cite{Shrock:1980ct}.  In case of $\tau$-leptons, the 2-body decays into hadrons $\tau \to NX$ are promising. If the hadronic system $X$ hadronizes into charged pions or kaons, then its mass and energy can be reconstructed at high precision. Using future $B$-factories with a large dataset of $\tau$ decays like $\tau^-\to N\pi^-\pi^+\pi^-$, stringent limits on the mixing parameter $|V_{\tau N}|^2$ can be placed~\cite{Kobach:2014hea}, as shown in Figure~\ref{fig:limVtauN}, where the (red, solid and dashed) contours labeled $B$-factory are the conservative and optimistic projected limits at 90\% C.L. from $\sim 10$ million $\tau$-decays. 

We should note here that the bounds from peak searches are very robust since they use only the kinematic features and minimal assumptions regarding the decay modes of the heavy neutrino. Moreover, since the heavy neutrino is assumed to be produced on-shell, these limits are valid irrespective of whether the heavy neutrino is a Majorana or Dirac particle.  

\subsubsection{Beam Dump Experiments}\label{sec:dump}
Another way to constrain the sterile neutrinos is via searches of their decay products. The sterile neutrinos are unstable due to their mixing with active neutrinos, and their decay rate is proportional to the mixing parameter $|V_{\ell N}|^2$. Thus, if kinematically allowed, they can be produced in semileptonic meson decays, and then subsequently decay into visible final states such as charged leptons, pions and kaons. These visible products can be searched for in beam dump experiments by placing the detector some distance away from the production site. The current 90\% C.L. limits from some of these beam dump experiments, such as PS191~\cite{Bernardi:1987ek}, NA3~\cite{Badier:1985wg}, CHARM~\cite{CHARM:1985, Vilain:1994vg, Orloff:2002de}, IHEP-JINR~\cite{Baranov:1992vq}, BEBC~\cite{CooperSarkar:1985nh}, FMMF~\cite{Gallas:1994xp}, NuTeV~\cite{Vaitaitis:1999wq} and NOMAD~\cite{Astier:2001ck} are shown in Figures~\ref{fig:limVeN}-\ref{fig:limVtauN}. It may be noted here that the PS191~\cite{Bernardi:1987ek} and CHARM~\cite{CHARM:1985} limits shown here assume that the sterile neutrinos interact only via CC. Including the NC interactions requires a reanalysis of the data which, in the context of $\nu$SM, gives twice stronger limits~\cite{Ruchayskiy:2011aa}. 

The proposed LBNE (now DUNE) experiment with a near detector could probe much smaller values of the mixing~\cite{Adams:2013qkq}, as illustrated in Figures~\ref{fig:limVeN}-\ref{fig:limVtauN} by the projected limits labeled `LBNE'. Here the heavy neutrinos are produced in charmed meson decays, and a near detector length of 30 m with a $\sim 5$ year exposure of $5\times 10^{21}$ protons on target is assumed. 
We have obtained the limits on individual mixing parameters from the corresponding limit on the sum of mixing $\sum_\ell |V_{\ell N}|^2$ assuming a normal hierarchy of light neutrinos.

\subsubsection{Rare LNV Decays of Mesons}\label{sec:rare_decay}
For heavy Majorana neutrinos, one could also look for rare LNV decays of mesons: $X_1^\pm \to \ell^\pm N$, $N\to \ell^\pm X_2^\mp$, which are forbidden in the SM. Searches for such decay modes have been performed in many experiments, such as CLEO, Belle, BaBar and LHCb~\cite{Agashe:2014kda}. The most stringent constraints come from $K^+\to \ell^+\ell^+\pi^-$ mode~\cite{Atre:2009rg}, as shown in Figures~\ref{fig:limVeN} and \ref{fig:limVmuN}. Here a realistic detector size of 10 m has been assumed. The corresponding limits from $D$ and $B$ meson decays~\cite{Castro:2013jsn, Yuan:2013yba, Wang:2014lda} are found to be weaker than the existing limits in the relevant mass region, and therefore, are not shown here, except the recent LHCb bound~\cite{Aaij:2014aba} in Figure~\ref{fig:limVmuN}, which was obtained using the 
$B^-\to \pi^+ \mu^- \mu^-$ decay mode with 3 fb$^{-1}$ of integrated luminosity collected at $\sqrt s=7$ and 8 TeV LHC.  

The bounds discussed in Sections~\ref{sec:dump} and \ref{sec:rare_decay} are less robust than those discussed in Section~\ref{sec:peak} because they are weakened, or even completely evaded, if the sterile neutrinos have other dominant decay modes into invisible particles. If the sterile neutrino decay length is shorter than the detector size, the number of signal events is suppressed by $|V_{\ell N}|^2$. On the other hand, if their decay length is larger than the detector size, the sterile neutrinos decay mostly outside the detector and the number of events is further suppressed by $|V_{\ell N}|^4$. This limitation could be overcome by increasing the flux of initial hadrons, e.g. in proposed fixed-target experiments such as SHiP~\cite{Anelli:2015pba} using high-intensity proton beams at the CERN SPS. The huge background due to multiparticle production inherent in hadron scatterings can be absorbed by adopting appropriate beam-dump techniques, thus allowing the sterile neutrinos to freely propagate into a decay volume. This experiment will improve the mixing sensitivity by up to four orders of magnitude~\cite{Alekhin:2015byh}, as shown by the projected limits in Figures~\ref{fig:limVeN}-\ref{fig:limVtauN}.

\subsubsection{$Z$-decays}\label{sec:zdecay}
For $M_N<M_Z$, using the possible production of heavy neutrinos in the $Z$-boson decay $Z\to \nu_{L\ell} N$ or $Z\to \bar{\nu}_{L\ell}N$~\cite{Dittmar:1989yg}, and its subsequent CC and NC decays, 95\% C.L. limits on the mixing parameters $|V_{\ell N}|^2$ were obtained by L3~\cite{Adriani:1992pq} and DELPHI~\cite{Abreu:1996pa} collaborations from a reanalysis of the LEP data. These limits are shown by the contours labeled `L3' (pink, dashed) in Figures~\ref{fig:limVeN} and \ref{fig:limVmuN}, and by the contours labeled `DELPHI' (dark green, dashed) in Figures~\ref{fig:limVeN}-\ref{fig:limVtauN}. 

A future high-luminosity $Z$-factory, such as the proposed FCC-ee experiment, will dramatically improve the sensitivity down to $|V_{\ell N}|^2\sim 10^{-12}$ for mixing with all neutrino flavors and covering a large phase space for heavy neutrino masses in the 10-80 GeV range~\cite{Blondel:2014bra, Abada:2014cca}. This is shown in Figures~\ref{fig:limVeN}-\ref{fig:limVtauN}, where the limits on individual mixing parameters are derived from the corresponding limit on the sum of mixing $\sum_\ell |V_{\ell N}|^2$~\cite{Blondel:2014bra} assuming a normal hierarchy of light neutrinos. We have also assumed $10^{12}$ $Z$-boson decays occurring between 10-100 cm from the interaction point. Increasing the number of $Z$-bosons and/or the range of decay length could further enhance these sensitivity limits, eventually reaching the theoretical expectation from the canonical seesaw formula, as shown by the (brown) dashed line labeled `Seesaw' in Figures~\ref{fig:limVeN}-\ref{fig:limVtauN}.

\subsubsection{Electroweak Precision Tests} \label{sec:ewpd}
Due to their mixing with active neutrinos, heavy neutrinos can affect various EW precision observables, such as the $Z$ invisible decay width and electroweak parameters in the SM~\cite{Nardi:1994iv, Nardi:1994nw, bergmann:1998rg}. The same mixing effects also show up in  non-unitarity of the leptonic mixing matrix~\cite{Antusch:2006vwa, abada:2007ux, Antusch:2014woa} and the violation of lepton universality in leptonic and semileptonic decays of pseudoscalar mesons~\cite{Abada:2012mc, Abada:2013aba, Asaka:2014kia}.   Using global fits to the EWPD, stringent model-independent constraints on $|V_{\ell N}|^2$ have been derived~\cite{delAguila:2008pw, Akhmedov:2013hec, Basso:2013jka, Blas:2013ana, Antusch:2015mia}. The current 90\% C.L. limits are 
are shown in Figures~\ref{fig:limVeN}-\ref{fig:limVtauN} (brown, dotted contours labeled `EWPD'). 
These limits are independent of the heavy neutrino mass for $M_N>M_Z$, and there is a mild mass dependence for lower $M_N$ values. Here we have only included the electroweak precision observables and lepton universality observables in the fit. The LFV observables are not included here, since they are more sensitive to the details of the Yukawa structure in the underlying model.

\subsection{Direct Collider Searches}\label{sec:direct}

Heavy neutrinos with masses of the order of electroweak scale can be directly produced on-shell at colliders. Such a direct search was performed in $e^+e^-$ annihilation at LEP~\cite{Acciarri:1999qj, Achard:2001qv}, assuming a single heavy neutrino production via its mixing with active neutrinos: $e^+e^-\to N\nu_{L\ell}$, followed by its decay via NC or CC interaction to the SM $W$, $Z$ or Higgs ($H$) boson: $N\to \ell W,~\nu_{L\ell} Z,~\nu_{L\ell} H$. Concentrating on the decay channel $N\to eW$ with $W\to$ jets, which would lead to a single isolated electron plus hadronic jets, the L3 collaboration put a 95\% C.L. upper limit on the mixing parameter $|V_{eN}|^2$ in a heavy neutrino mass range between 80 and 205 GeV~\cite{Achard:2001qv}, as shown by the (red, solid) contour labeled `L3 II' in Figure~\ref{fig:limVeN}. This search was mainly limited by the maximum center-of-mass energy $\sqrt s=208$ GeV at LEP. Future lepton colliders can significantly improve the sensitivity in this mass region, as illustrated in Figure~\ref{fig:limVeN} by the projected limit labeled `ILC', which is obtained assuming a $\sqrt s=500$ GeV ILC with luminosity of 500 fb$^{-1}$~\cite{Banerjee:2015gca}.

In the context of hadron colliders, a Majorana heavy neutrino leads to the smoking gun lepton-number violating signature of same-sign dilepton plus jets with no missing transverse energy: $pp~(p\bar{p}) \to W^* \to N\ell^\pm \to \ell^\pm \ell^\pm jj$~\cite{Keung:1983uu, Pilaftsis:1991ug, Datta:1993nm, Almeida:2000pz, Panella:2001wq, Han:2006ip, Bray:2007ru, delAguila:2007em, Atre:2009rg}. An inclusive search for new physics with same-sign dilepton signals was first performed in $p\bar{p}$ collisions at the Tevatron~\cite{Abulencia:2007rd}. After the inauguration of the LHC era, the CMS and ATLAS collaborations have performed direct {\em exclusive} searches for the on-shell production of heavy neutrinos above the $Z$-threshold. The previous searches with 4.7 fb$^{-1}$ data at $\sqrt s=7$ TeV LHC set 95\% C.L. limits on $|V_{\ell N}|^2\lesssim 10^{-2}-10^{-1}$ (with $\ell=e, \mu$) for heavy neutrino masses up to 300 GeV~\cite{Chatrchyan:2012fla, ATLAS:2012yoa}. More recently, these limits were extended for masses up to 500 GeV with 20 fb$^{-1}$ data at $\sqrt s=8$ TeV~\cite{Khachatryan:2015gha, klinger:2014}, and are shown in Figures~\ref{fig:limVeN} (ATLAS) and \ref{fig:limVmuN} (ATLAS and CMS). For $M_N\sim 100$ GeV, the direct limits in the muon sector are comparable to the indirect limits on $|V_{\mu N}|^2\lesssim 10^{-3}$ imposed by the electroweak precision data~\cite{delAguila:2008pw, Blas:2013ana} and LHC Higgs data~\cite{BhupalDev:2012zg, Cely:2012bz}. With the run-II phase of the LHC starting later this year with more energy and  higher luminosity, the direct search limits could be extended for heavy neutrino masses up to a TeV or so. Also note that the LFV processes put stringent constraints on the product $|V_{\ell N}V^*_{\ell' N}|$ (with $\ell\neq \ell'$)~\cite{ilakovac:1995kj, Ilakovac:1995km, Dinh:2012bp, Alonso:2012ji}, but do not restrict the individual mixing parameters $|V_{\ell N}|^2$ in a model-independent way. So the direct searches provide a complementary way to probe the light-heavy neutrino mixing in the seesaw paradigm. 

\begin{figure}[t]
\centering
\includegraphics[width=5cm]{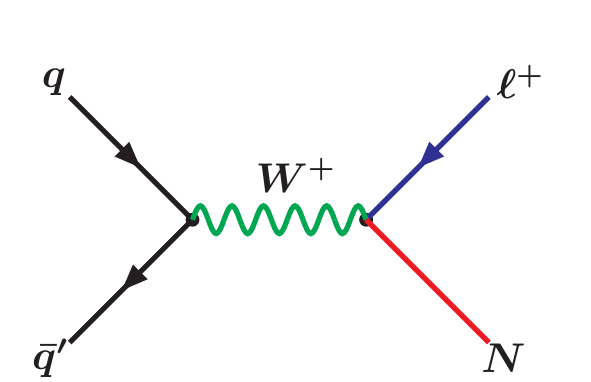}
\caption{Feynman diagram for heavy neutrino production at the LHC via the $s$-channel Drell-Yan process.}
\label{fig:prodNs} 
\end{figure}

\begin{figure}[t]
\centering
\includegraphics[width=5cm]{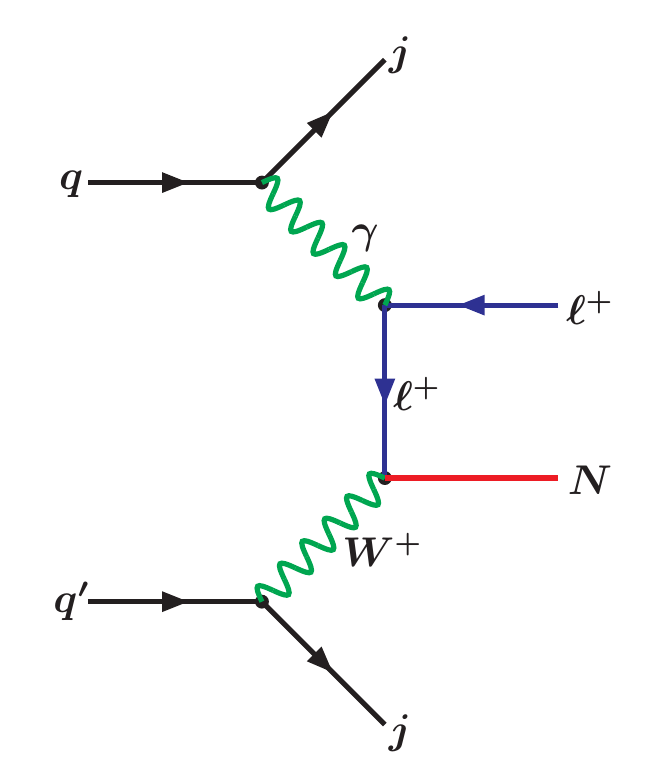} 
\hspace{0.5cm} 
\includegraphics[width=5.5cm]{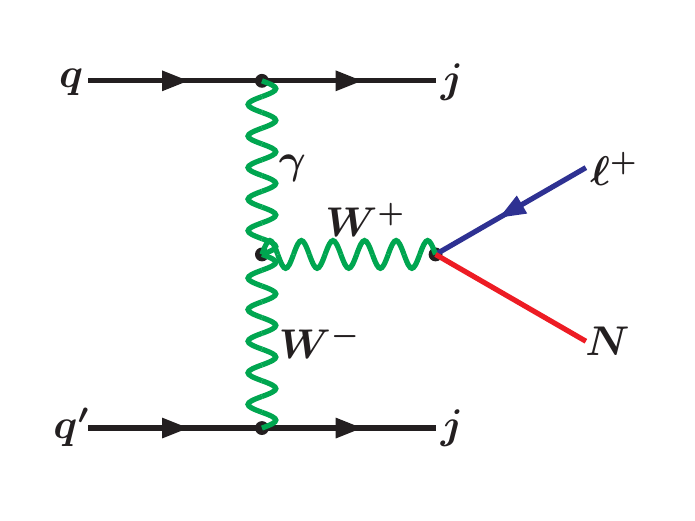}  \\
\includegraphics[width=6.5cm]{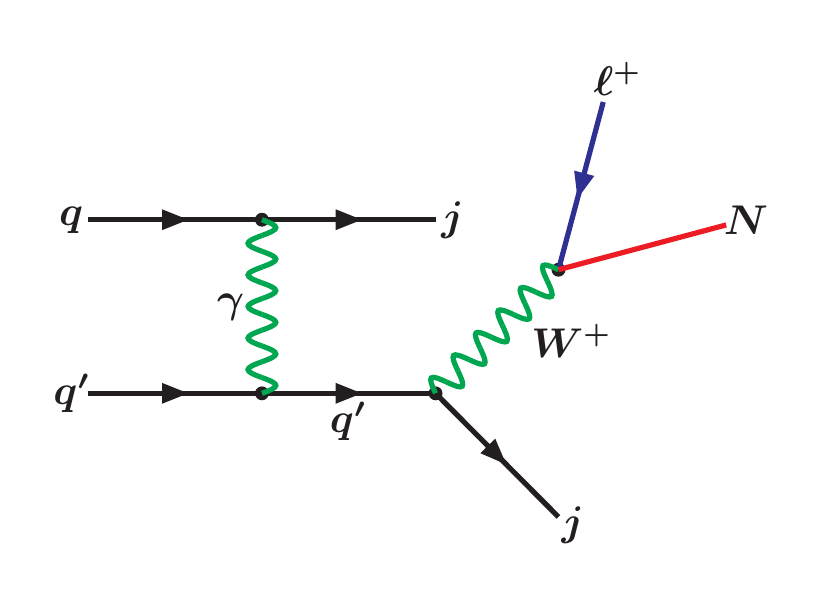} 
\hspace{0.2cm} 
\includegraphics[width=5.5cm]{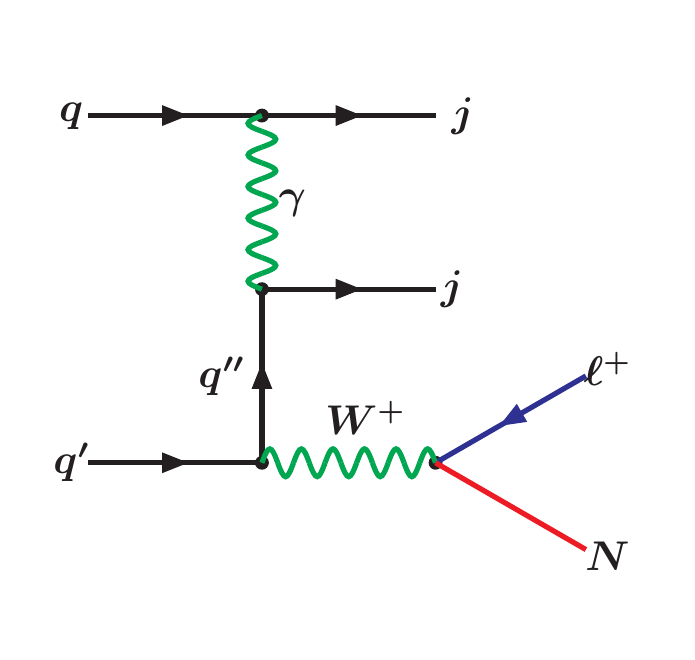} 
\caption{Feynman diagrams for heavy neutrino production at the LHC via the $t$-channel photon-mediated processes~\cite{Dev:2013wba}.}
\label{fig:prodNt} 
\end{figure}

\begin{figure}[t!]
\centering
\includegraphics[width=10cm]{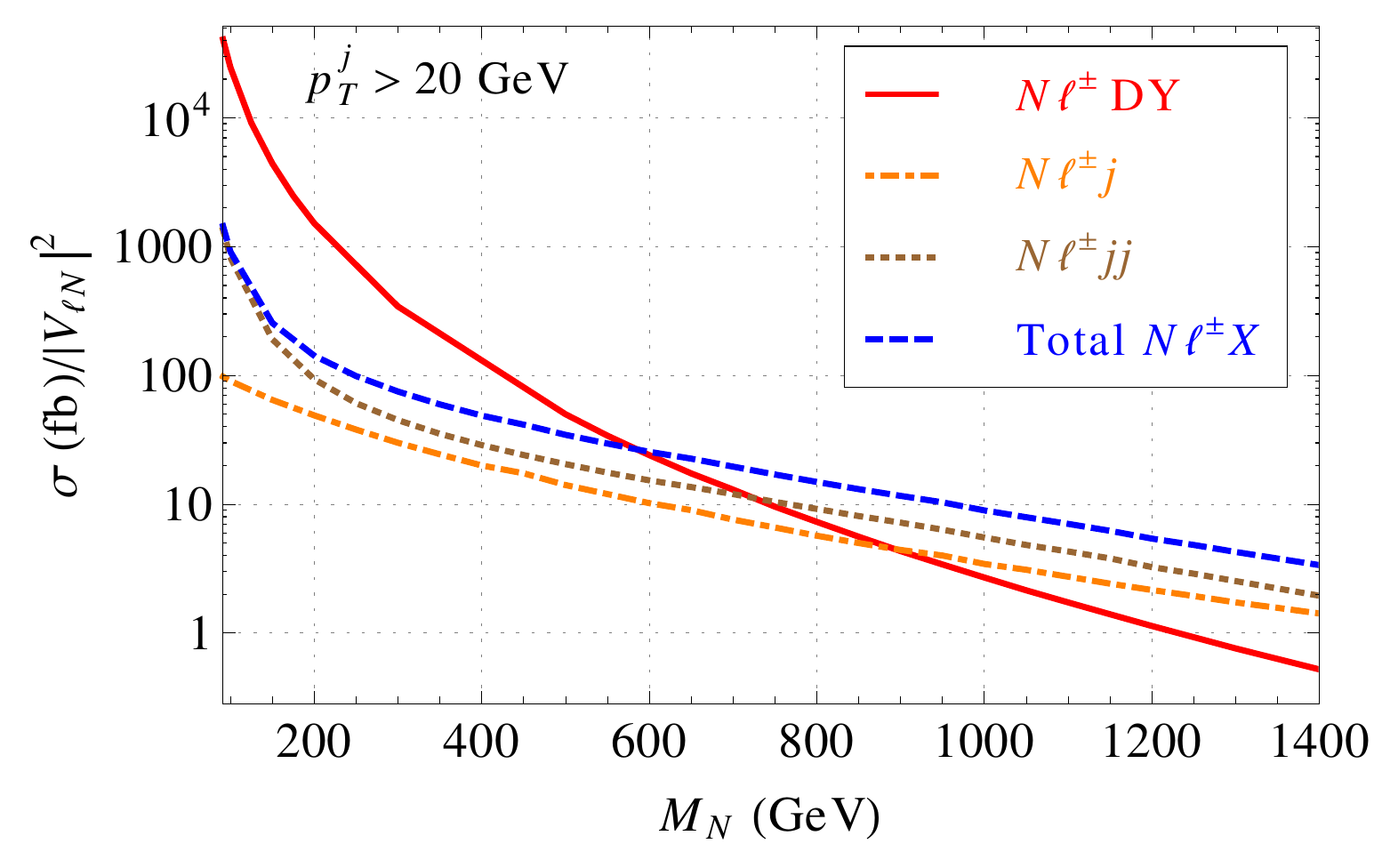}
\caption{Comparison of the cross sections for heavy neutrino production at $\sqrt s=14$ TeV LHC via the $s$-channel (Figure~\ref{fig:prodNs}) and $t$-channel (Figure~\ref{fig:prodNt}) diagrams.}
\label{fig:diagrampt} 
\end{figure}

All the direct searches at the LHC so far have only considered the simplest production process for heavy neutrinos through an $s$-channel $W$-exchange~\cite{Pilaftsis:1991ug, Datta:1993nm, Almeida:2000pz, Panella:2001wq, Han:2006ip, Bray:2007ru, delAguila:2007em, Atre:2009rg}, as shown in Figure~\ref{fig:prodNs}. However, there exists another collinearly enhanced electroweak production mode involving $t$-channel exchange of photons: $pp\to W^*\gamma^* \to N\ell^\pm jj$~\cite{Dev:2013wba}, cf. Figure~\ref{fig:prodNt}, which gives a dominant contribution to the heavy neutrino production cross section for higher $M_N$ values. This is mainly because of the fact that with increasing heavy neutrino mass, the production cross section for the $t$-channel process drops at a rate slower than that of the $s$-channel process, though the exact cross-over point depends crucially on the selection cut for the $p_T$ of the additional jet associated with the virtuality of the $t$-channel photon~\cite{Dev:2013wba, Alva:2014gxa}. The photon-mediated process $pp\to W^*\gamma^* \to N\ell^\pm jj$ has two contributions: an inelastic part with $t$-channel virtual photon and an elastic part $p\gamma\to N\ell^\pm j$ with a real photon emission from one of the protons. The elastic part is calculated using an effective photon structure function for the proton~\cite{Budnev:1974de, Martin:2014nqa}, whereas the inelastic part is computed for a non-zero minimum $p_T$ of the jet associated with the virtuality of the photon~\cite{Dev:2013wba}.  A comparison of the production cross sections for the $s$-channel Drell-Yan (DY) process $pp\to W^* \to N\ell^\pm$  and the photon initiated processes is made in Figure~\ref{fig:diagrampt} for a representative value of $p_{T,{\rm min}}^j=20$ GeV which, alongwith a jet separation cut $\Delta R^{jj}>0.4$, is sufficient to ensure that the photon-mediated processes are collinear safe. We find that the total photon-initiated contribution becomes dominant over the DY cross section for $M_N\gtrsim 600$ GeV. Here, we have not included the QCD corrections, which could further lower the cross-over point to the level presented in~\cite{Dev:2013wba}, but this requires a more careful analysis and will be presented elsewhere. Note that the numerical results shown in Figure~\ref{fig:diagrampt} are slightly different from those presented in~\cite{Alva:2014gxa}, which can be mainly attributed to the choice of regulator used to treat the collinear behavior.  

In any case, including the collinear enhancement effect could further enhance the heavy neutrino signal sensitivity at the next run of the LHC~\cite{Dev:2013wba}. As an illustration, we have shown in Figures~\ref{fig:limVeN} and \ref{fig:limVmuN} projected conservative limits with 300 fb$^{-1}$ data at $\sqrt s=14$ TeV (blue, dashed contours labeled `LHC 14'), assuming that the cross-section limits are at least as good as the existing ones at $\sqrt s=8$ TeV, as reported in~\cite{klinger:2014}.  The direct collider limits for $M_N<100$ GeV are not likely to improve significantly with higher collision energy, due to the increased pile-up effects, thus obfuscating the low-$p_T$ leptons produced by the decay of a low-mass heavy neutrino. Instead, a displaced vertex search might be useful to probe the low-mass range between 3-80 GeV for mixing values $10^{-7}\lesssim |V_{\ell N}|^2\lesssim 10^{-5}$~\cite{Helo:2013esa}. 

For heavy Dirac neutrinos as predicted in theories with approximate $L$-conservation [cf.~\eqref{lag_inv}], the same-sign dilepton signal is suppressed. In this case, the golden channel is the trilepton channel: $pp\to W^*\to N\ell^\pm \to \ell^\pm \ell^\mp \ell^\pm + \slashed{E}_T$~\cite{delAguila:2008cj, delAguila:2008hw, delAguila:2009bb, Chen:2011hc, Das:2012ze, Das:2014jxa, Bambhaniya:2014kga}. Using this trilepton mode and also taking into account the infrared enhancement effects~\cite{Dev:2013wba}, direct limits on the mixing of heavy Dirac neutrinos with electron and muon neutrinos were obtained~\cite{Das:2014jxa} by analyzing the tri-lepton data from $\sqrt s=8$ TeV LHC~\cite{Chatrchyan:2014aea}. 

Finally, we note that there exist no direct collider searches for heavy neutrinos involving tau-lepton final states. This is mainly due to the experimental challenges of $\tau$ reconstruction at a hadron collider. The situation is expected to improve in future with better $\tau$-tagging algorithms and/or in cleaner environments of a lepton collider. 

\section{Heavy Triplets at Colliders} \label{sec:3}
Unlike the minimal Type-I seesaw messengers which, being SM gauge singlets, can only communicate with the SM sector through their mixing with the active neutrinos, the Type-II and III seesaw messengers are $SU(2)_L$ triplet scalar $(\Delta^{++},\Delta^+,\Delta^0)$ and fermion $(\Sigma^{+},\Sigma^0,\Sigma^-)$ fields respectively, and hence, can be {\em directly} produced at the LHC via their gauge interactions.  For Type-II seesaw~\cite{Schechter:1980gr, Magg:1980ut, Cheng:1980qt, lazarides:1980nt, mohapatra:1981yp}, the smoking gun signal would be the detection of a doubly-charged scalar with LNV interactions. For this scenario, the most relevant production channels at the LHC are $pp\to Z^*/\gamma^* \to \Delta^{++}\Delta^{--}, \Delta^+\Delta^-$, $pp\to W^{\pm *}W^{\pm *}\to \Delta^\pm \Delta^\pm$ and $pp\to W^* \to \Delta^{\pm\pm}\Delta^\mp, \Delta^{\pm\pm}W^\mp$~\cite{akeroyd:2005gt, han:2007bk, akeroyd:2007zv, perez:2008ha, Akeroyd:2009hb, Akeroyd:2010ip, delAguila:2008cj, Melfo:2011nx, Sugiyama:2012yw, delAguila:2013mia, Chen:2014qda, Han:2015hba}.  The doubly-charged scalar boson has the following possible decay channels: $\ell^\pm\ell^\pm$, $W^\pm W^\pm$, $W^\pm \Delta^\pm$ and $\Delta^\pm\Delta^\pm$, if kinematically allowed. For the triplet VEV $v_\Delta\lesssim 0.1~{\rm MeV}$, the doubly charged Higgs couplings to $W^\pm$ is suppressed and for a nearly degenerate triplet mass spectrum, the dominant decay mode of $\Delta^{\pm\pm}$ is same-sign dileptons~\cite{Melfo:2011nx, Chun:2013vma}. In this case, the current 95\% C.L. experimental lower bound on the doubly-charged triplet scalar mass is between 490-550 GeV, depending on the final lepton flavor~\cite{Chatrchyan:2012ya, ATLAS:2012hi, ATLAS:2014kca}. For $v_\Delta\gtrsim 0.1$ MeV, the Yukawa couplings of the $\Delta$ fields to leptons is suppressed and $\Delta^{\pm\pm}$ dominantly decays to same-sign dibosons, in which case the collider limits are significantly weaker~\cite{Chiang:2012dk, Kanemura:2013vxa, kang:2014jia, Kanemura:2014goa, Khachatryan:2014sta}. These mass bounds could be significantly improved in the upcoming run II phase of the LHC. A future lepton collider such as the ILC will offer an almost background-free environment for the doubly-charged scalar detection, if kinematically allowed, through the single production process $e^+e^-\to e^\pm \ell^\pm \Delta^{\mp\mp}$~\cite{Barenboim:1996pt} or pair-production process $e^+e^-\to \Delta^{++}\Delta^{--}$~\cite{Yagyu:2014aaa}. We note here that a relatively light charged scalar will affect the loop-induced decays of the SM Higgs boson $h\to \gamma\gamma$~\cite{Melfo:2011nx, Arhrib:2011vc, Akeroyd:2012ms, Carena:2012xa, Dev:2013ff, Chabab:2014ara, Arhrib:2014nya} and $h\to Z\gamma$~\cite{Carena:2012xa, Dev:2013ff, Chen:2013dh, Chabab:2014ara, Arhrib:2014nya}. In fact, for a given enhancement in these decay rates over the SM prediction, one can set an {\em upper} limit on the charged scalar mass in the minimal Type-II seesaw model using vacuum stability and perturbative arguments~\cite{Dev:2013ff}.

For Type-III seesaw~\cite{foot:1988aq}, the relevant production mechanisms at the LHC are $pp\to Z^*/\gamma^*\to \Sigma^+\Sigma^-$, $pp\to Z^*\to \ell^\pm \Sigma^\pm$ and $pp\to W^{\pm *}\to \Sigma^\pm \Sigma^0, \ell^\pm \Sigma^0$~\cite{franceschini:2008pz, delAguila:2008cj, Arhrib:2009mz, Li:2009mw}. The neutral component $\Sigma^0$ has the same decay modes as the heavy sterile neutrino discussed in Section~\ref{sec:2.2}, i.e. $\Sigma^0\to \ell^\pm W^\mp,\nu_\ell(\bar\nu_\ell) Z, \nu_\ell(\bar\nu_\ell) H$, whereas the charged fermion has the LNV decay modes $\nu_\ell(\bar\nu_\ell)W^\mp,\ell^\pm Z, \ell^\pm H$.  The current  95\% C.L. experimental lower bound on the fermion triplet mass is 245 GeV~\cite{CMS:2012ra, ATLAS:2013hma}. This model has promising discovery prospects at the ILC as well~\cite{Yue:2010zzb, Garg:2014jva}.

\section{Extended Gauge Sectors} \label{sec:4}
Within the SM gauge group, there is no explanation for the origin of the Majorana masses of the seesaw messengers. This problem can be solved by extending the SM gauge group so that the Majorana mass can be associated with the spontaneous breaking of the extra gauge symmetry. The simplest of such additional gauge symmetries is $U(1)_{B-L}$ whose breaking could set the mass scale of the RH neutrinos in the Type-I seesaw. A natural embedding of the RH Majorana neutrinos as well as the scalar triplets can be found in the LRSM based on the gauge group $SU(2)_L\times SU(2)_R\times U(1)_{B-L}$.  Therefore, we will focus on these two additional gauge symmetries in the following.

\subsection{Additional $U(1)$}\label{sec:4.1}

The simplest extension of the SM gauge group to explain the heavy Majorana neutrino mass in \eqref{lagM} is the inclusion of an additional $U(1)$ gauge group along with an associated $Z'$ gauge boson. This symmetry can be added by hand to the SM but it could also naturally arise from a UV-complete theory such as the Pati-Salam model, $SO(10)$ or $E_6$ GUTs. For a review of various $Z'$ models, see e.g.~\cite{Langacker:2008yv}. The mass and couplings of the $Z'$ boson are strongly constrained by EWPD. Constraints from lepton universality at the $Z$ peak puts a lower limit of $M_{Z'} \gtrsim \mathcal{O}(1)$~TeV~\cite{Polak:1991pc}, whereas direct searches at the LHC exclude $M_{Z'}$ below about 2 TeV~\cite{Agashe:2014kda}. Similarly, the mixing angle between $Z'$ and the SM $Z$ is limited to be less than $\mathcal{O}(10^{-4})$. 

\begin{figure}[t!]
\centering
\includegraphics[clip,width=0.45\textwidth]{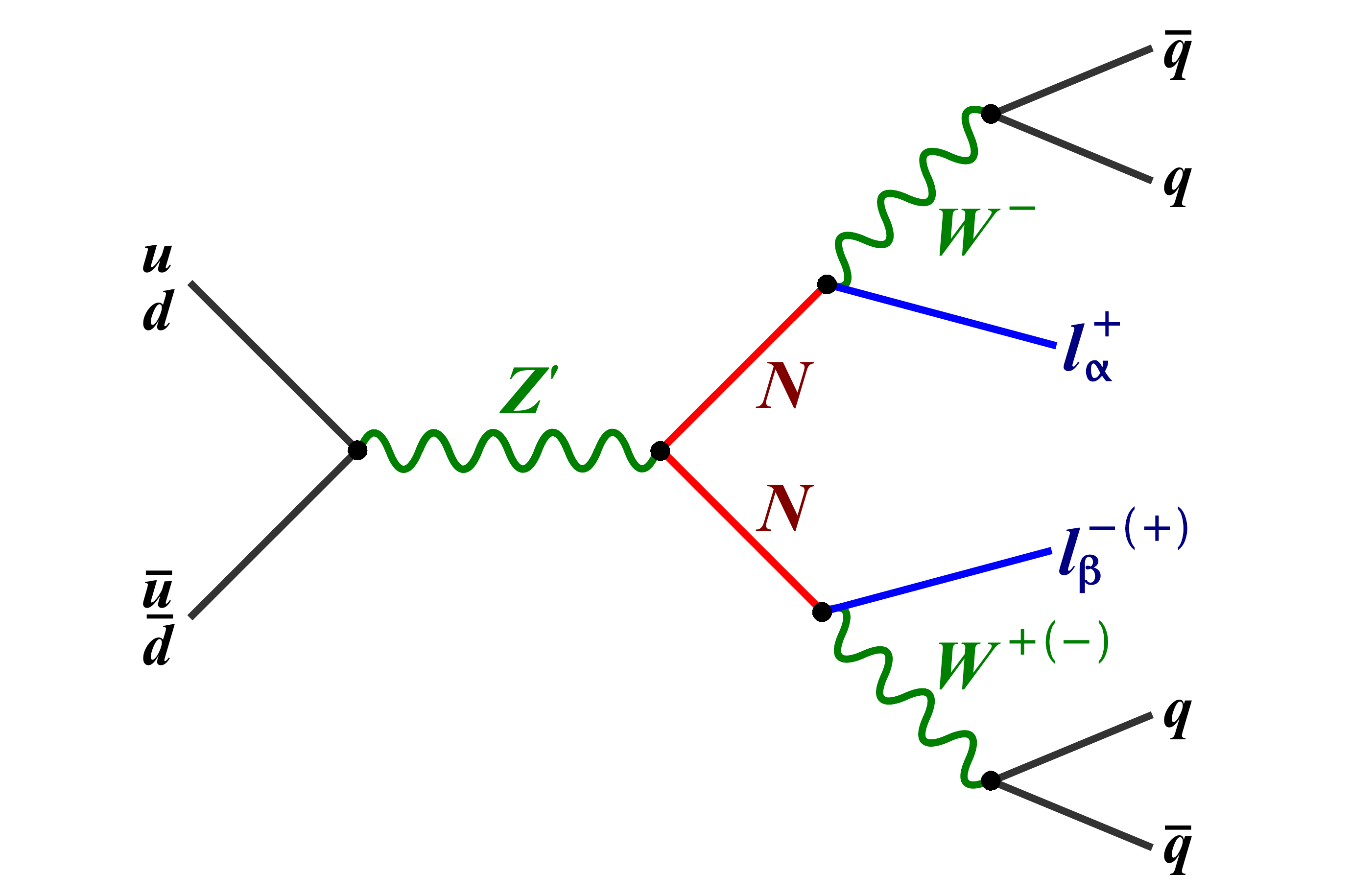}
\includegraphics[clip,width=0.45\textwidth]{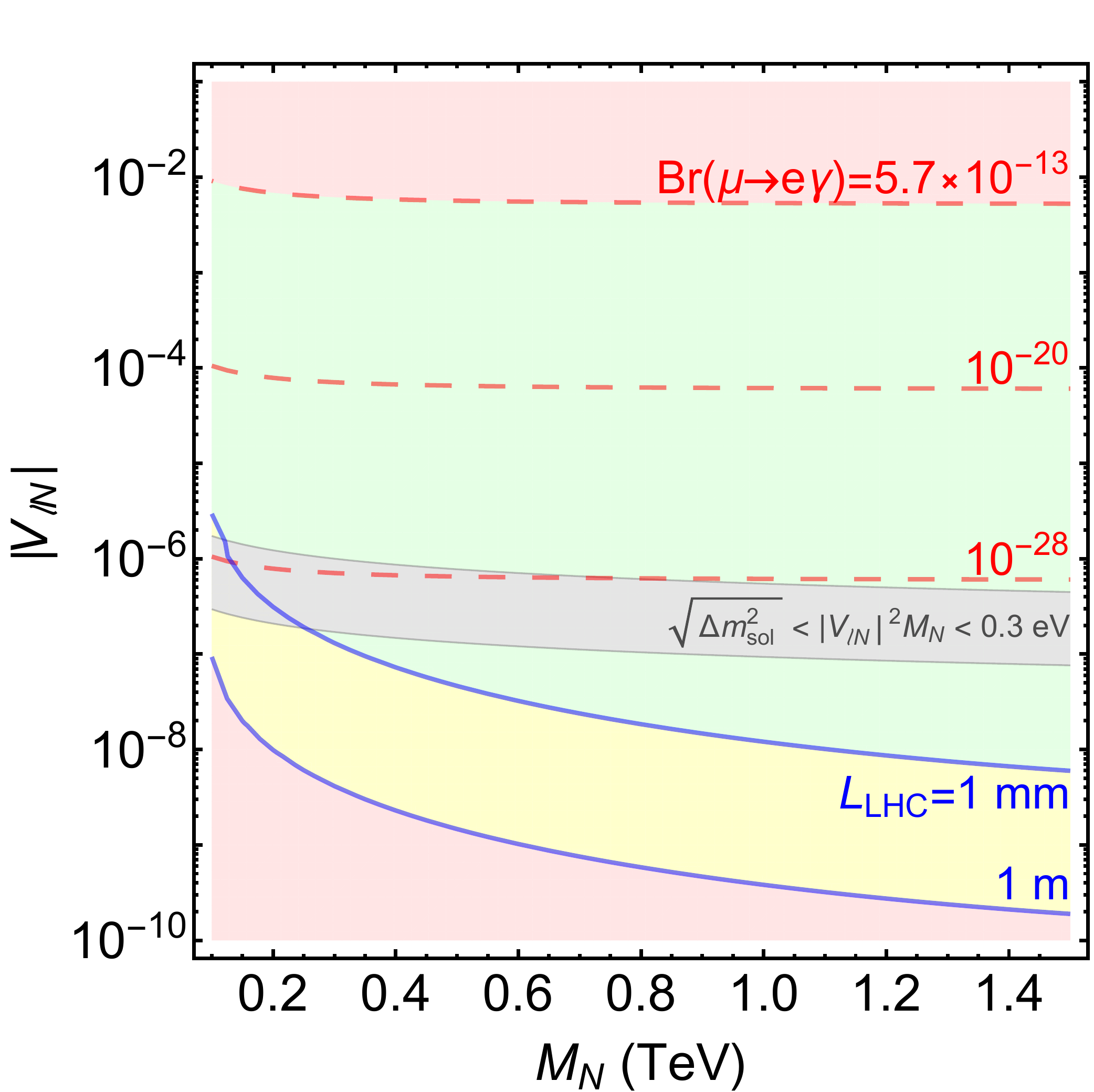}
\caption{{\em Left:} Feynman diagram for heavy neutrino production via the $Z'$ resonance. {\rm Right:} Heavy neutrino decay length as a function of its mass $M_N$ and mixing $V_{\ell N}$ (solid blue contours). The dashed red contours denote Br$(\mu\to e\gamma)$, with the shaded red region on top excluded by the current MEG limit~\cite{Adam:2013mnn}. The gray shaded band highlights the parameter range where light neutrino mass scales between $\sqrt{\Delta m^2_\text{sol}}$ and 0.3~eV are generated within the canonical type-I seesaw mechanism.}
\label{fig:zprime_diagram} 
\end{figure}

With regard to heavy neutrinos, the main phenomenological advantage is the possibility that the heavy neutrinos are charged under the additional $U(1)$. This would provide a new and potentially strong production channel at colliders. A Feynman diagram for resonant heavy neutrino production at the LHC via such a $Z'$ portal with a final state of two leptons and four jets~\cite{Perez:2009mu} is shown in Figure~\ref{fig:zprime_diagram} (left panel). An important point to note is that the total cross section of this process is {\em independent} of the mixing strength $V_{\ell N}$. This process could therefore be observed at the LHC as long as the total decay width of the heavy neutrino is large enough such that it decays within the detector. The relevant parameter range for this to occur is shown in Figure~\ref{fig:zprime_diagram} (right panel)~\cite{Deppisch:2013cya}. Even the canonical Type-I seesaw with small mixing $|V_{\ell N}| \lesssim 10^{-6}$ for TeV-scale $M_N$ can be potentially probed in this case, possibly through displaced vertices. This also includes the potential to observe LFV signatures, despite the unobservably small LFV rates for low-energy processes such as $\mu\to e\gamma$, as they are strongly suppressed by such a small mixing. 

%

\subsection{Left-Right Symmetric Models}
\label{sec:4.2}

The minimal LRSM which extends the SM gauge symmetry to $SU(2)_L\times SU(2)_R\times U(1)_{B-L}$~\cite{Pati:1974yy, Mohapatra:1974gc, Senjanovic:1975rk} provides a simple UV-complete seesaw model, where the key ingredients of seesaw, i.e. the RH neutrinos and their Majorana masses, appear naturally. The presence of RH neutrinos is a necessary ingredient for the restoration of Left-Right symmetry and is also required by anomaly cancellation, whereas the seesaw scale is identified as the breaking of the $SU(2)_R$ symmetry. It is worth noting that in the presence of three RH neutrinos, the $(B-L)$-symmetry which was a global symmetry in the SM becomes a gauge symmetry in the LRSM, as the gauge anomalies cancel by satisfying the condition Tr$(B-L)^3=0$. Moreover, the electric charge formula takes a form similar to the Gell-Mann - Nishijima relation: $Q=I_{3L}+I_{3R}+(B-L)/2$, where $I_{3L}$ and $I_{3R}$ are the third components of isospin under $SU(2)_L$ and $SU(2)_R$ respectively, and the SM hypercharge can now be understood as $Y/2= I_{3R}+(B-L)/2$~\cite{Marshak:1979fm, mohapatra:1980qe}.  

In the LRSM, leptons are assigned to the multiplets $L_\ell = (\nu_\ell, \ell)_L$ and $R_\ell = (N_\ell, \ell)_R$ (where $\ell=e,\mu,\tau$ is the generation index) with the quantum numbers $(\mathbf{2}, \mathbf{1}, -1)$ and $(\mathbf{1}, \mathbf{2}, -1)$ respectively under $SU(2) \times SU(2) \times U(1)_{B-L}$. The Higgs sector of the minimal LRSM contains a bidoublet $\phi$ with quantum numbers $(\mathbf{2},\mathbf{2},0)$ and two triplets ${\Delta}_{L,R}$ with quantum numbers $(\mathbf{3},\mathbf{1},2)$ and $(\mathbf{1},\mathbf{3},2)$ respectively. The VEV $v_R$ of the neutral component of ${\Delta}_R$ breaks the gauge symmetry $SU(2)_R\times U(1)_{B-L}$ to $U(1)_Y$ and gives masses to the RH gauge bosons $W_R$, $Z_R$ boson and the RH neutrinos $N_R$. The VEVs $(\kappa,\kappa')$ of the neutral components of the bidoublet $\phi$ break the SM symmetry and are therefore of the order of the electroweak scale. 

The LRSM Lagrangian relevant for the neutrino mass is given by 
\begin{align}
-{\cal L}_Y \ = \ h \bar{L}\phi R + \tilde{h}\bar{L}\tilde{\phi} R + f_L L^{\sf T}Ci\sigma_2 \Delta_L L+ f_R R^{\sf T}Ci\sigma_2 \Delta_R R + {\rm H.c.}\; ,
\label{lagLR}
\end{align}
where $\tilde{\phi}=\sigma_2 \phi^* \sigma_2$ and $h,\tilde{h}, f_{L,R}$ are $3\times 3$ complex Yukawa couplings. After symmetry breaking, \eqref{lagLR} leads to the Dirac mass matrix $M_D=h\kappa+\tilde{h}\kappa'$ and the Majorana mass matrices $M_L=f_Lv_L$ and $M_R=f_Rv_R$ for the light and heavy neutrinos respectively, where $v_L$ is the VEV of the neutral component of ${\Delta}_L$. This leads to the neutrino mass matrix
\begin{align}
	{\cal M}_\nu \ = \ 
	\begin{pmatrix}
		M_L   & M_D \\ 
		M_D^{\sf T} & M_R
	\end{pmatrix},
\label{eq:matr}
\end{align}
as compared to the Type-I case given by \eqref{eq:seesaw}. In the usual seesaw approximation $\|M_D\|\ll \|M_R\|$, diagonalizing \eqref{eq:matr} leads to the light neutrino mass matrix of the form
\begin{align}
	M_\nu \ = \ M_L-M_D M_R^{-1} M_D^{\sf T} \; ,
\label{Mnu-LR}
\end{align}
where the second term on the RHS is the Type-I seesaw contribution which is inversely proportional to $v_R$, whereas the first one is the Type-II seesaw contribution which is directly proportional to $v_L$. It should be noted here that in the minimal LRSM, if charge conjugation is the discrete L-R symmetry, $M_D=M_D^{\sf T}$ and $M_L=(v_L/v_R)M_R$. In this case, the Dirac Yukawa couplings are generically constrained by the light and heavy neutrino mass and mixing parameters~\cite{Nemevsek:2012iq}.  
However, there are exceptions, e.g. in $A_4$ symmetry-based models, where 
\begin{align}
M_D \ \propto \ \left(\begin{array}{ccc}
1 & \omega & \omega^2 \\
\omega & \omega^2 & 1 \\
\omega^2 & 1 & \omega
\end{array}\right) \; ,
\label{MD-A4}
\end{align} 
with $\omega^3=1$ and $M_N=m_N\mathbf{1}_3$. For such symmetry-based models, $M_\nu$ vanishes {\em identically} for $v_L = 0$, independently of the size of the Dirac Yukawa couplings.\footnote{If parity is the discrete L-R symmetry which implies $M_D=M_D^\dag$, a similar construction can be made by interchanging the second and third columns of $M_D$ in~\eqref{MD-A4}.} 
Likewise, in versions of LRSM where parity and $SU(2)_R$ gauge symmetry scales are decoupled~\cite{Chang:1983fu}, the Dirac Yukawa couplings could be sizable~\cite{Dev:2013oxa}, while being consistent with the light neutrino data.

Since processes induced by the RH currents and particles have not been observed so far, $v_R$ has to be sufficiently large. In particular, hadronic flavor changing neutral current effects restrict $M_{W_R} \gtrsim 3$ TeV~\cite{Beall:1981ze, Zhang:2007da, Maiezza:2010ic, Bertolini:2014sua}, assuming that the $SU(2)_R$ gauge coupling $g_R$ has the same strength as the $SU(2)_L$ gauge coupling $g_L$. Direct search limits from the $\sqrt s=7$ and 8 TeV LHC data put similar constraints on $M_{W_R}$, depending on the heavy neutrino mass~\cite{ATLAS:2012ak, Khachatryan:2014dka}. This translates into a lower limit on $v_R=\sqrt{2} M_{W_R}/g_R \gtrsim 6.5$ TeV. On the other hand, for the the left-triplet VEV $v_L$, the electroweak $\rho$-parameter constraints set an upper limit on $v_L \lesssim 5$ GeV~\cite{Kanemura:2012rs}. 

\begin{figure}[t!]
\centering
\includegraphics[clip,width=0.45\textwidth]{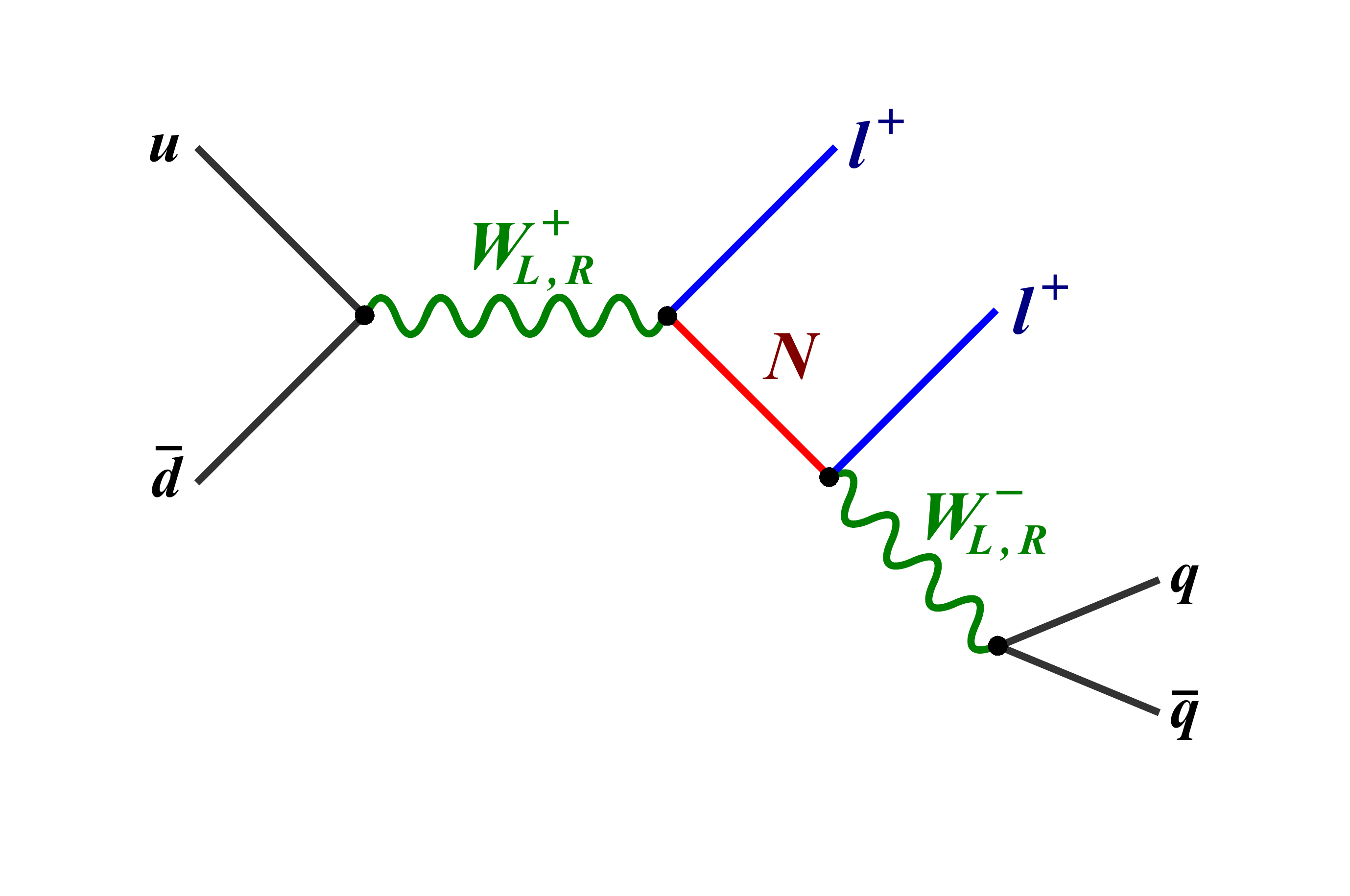}
\includegraphics[clip,width=0.45\textwidth]{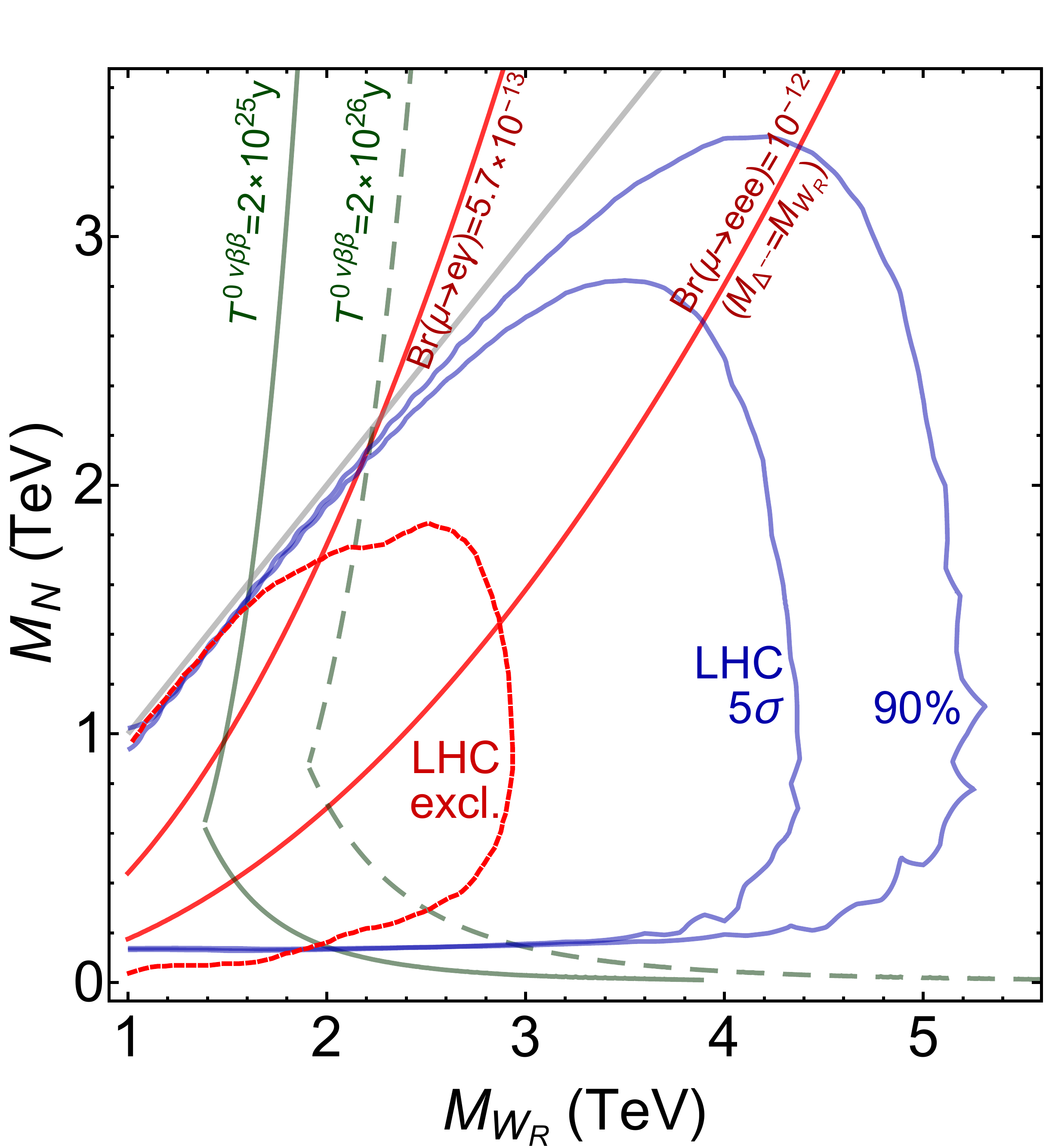}
\caption{{\em Left:} Feynman diagrams contributing to the `smoking gun' collider 
signal of LNV ($\ell^\pm\ell^\pm jj$) in the LRSM through the production via SM $W_{(L)}$ and heavy $W_R$, giving rise to 4 different contributions: RR, RL, LL, LR. {\em Right:} Comparison of LNV event rates via the RR diagram at the LHC and in $0\nu\beta\beta$ experiments~\cite{Das:2012ii}. The solid blue contours give the signal significance of $5\sigma$ and 90\% at the LHC with 14~TeV and $\mathcal{L}=300\text{ fb}^{-1}$. The area denoted `LHC excl.' is excluded by current LHC searches in the electron channel~\cite{Khachatryan:2014dka}. The green contours show the sensitivity of current and future $0\nu\beta\beta$ experiments, assuming dominant doubly-charged Higgs or heavy neutrino exchange and the red contours show the sensitivity of LFV processes as denoted.}
\label{fig:LR}
\end{figure}

Due to the presence of RH gauge interactions, the LRSM gives rise to a number of new contributions to both LNV and LFV processes; see e.g.~\cite{Das:2012ii, Deppisch:2012vj, Barry:2013xxa}.  In particular, there are several diagrams that contribute to the $0\nu\beta\beta$ amplitude:~(i) standard light neutrino exchange with mass helicity flip~\cite{Racah:1937qq, Furry:1939qr}, (ii) RH neutrino and RH gauge boson exchange~\cite{mohapatra:1979ia, mohapatra:1981yp}, (iii) RH Higgs triplet exchange~\cite{Mohapatra:1981pm}, and (iv) mixed LH-RH contributions~\cite{hirsch:1996qw, Parida:2012sq, Awasthi:2013ff, Barry:2013xxa, Huang:2013kma, Dev:2014xea}. The latter depend on the size of the left-right neutrino mixing $V_{\ell N} \simeq M_D M_R^{-1}$. In the Type-II seesaw dominance~\cite{Tello:2010am, Chakrabortty:2012mh}, neglecting the mixed left-right as well as the canonical light-neutrino contributions, the $0\nu\beta\beta$ half-life due to purely RH currents can be written as
\begin{align}
	\frac{1}{T_{1/2}^{0\nu}} \ = \ 
	{\cal A}\left|m_p\left(\frac{M_W}{M_{W_R}}\right)^4\sum_i \frac{V^2_{ei}}{M_i}\right|^2 \;, 
\label{half-R}
\end{align}
where $V_{ei}$ is the mixing matrix for the RH neutrinos $N_i$ with mass eigenvalues $M_i$ and ${\cal A}$ is defined below \eqref{half-life}. Using \eqref{half-R} and the current experimental limits on $T_{1/2}^{0\nu}$, lower limits on the RH gauge boson and RH neutrino masses can be derived~\cite{Das:2012ii, Dev:2013vxa}. This is illustrated in Figure~\ref{fig:LR}, where the excluded areas from $0\nu\beta\beta$ searches are shown. Similarly, low-energy LFV processes such $\mu \to e\gamma$ and $\mu \to 3e$ can be drastically enhanced in the LRSM, with a host of new contributions~\cite{Cirigliano:2004mv}. This is also illustrated in Figure~\ref{fig:LR} using maximal $e\mu$ flavor mixing of the heavy neutrinos~\cite{Das:2012ii}.

As for the LHC phenomenology, the presence of RH gauge interactions could lead to significant enhancement of the LFV/LNV signal. There are several contributions to the smoking gun LNV signal of same-dilepton plus two jets, as summarized in Figure~\ref{fig:LR}~(left). Even if the left-right neutrino mixing is small, heavy RH neutrinos could be directly produced via $s$-channel $W_R$ exchange and subsequently decay via the same $W_R$~\cite{Keung:1983uu}. The potential to discover LFV and LNV at the LHC in this scenario has been analyzed in~\cite{Nemevsek:2011hz, Chakrabortty:2012pp, AguilarSaavedra:2012fu, Das:2012ii, AguilarSaavedra:2012gf}. Figure~\ref{fig:LR} compares the sensitivity of LNV searches at the LHC with the sensitivity of $0\nu\beta\beta$ experiments. The $\sqrt s=14$~TeV LHC might be ultimately able to probe RH gauge boson masses up to $M_{W_R}=6$ TeV~\cite{ferrari:2000sp}, whereas a futuristic $\sqrt s=80\: (100)$ TeV hadron collider could probe up to $M_{W_R}=26.6\:(35.5)$ TeV~\cite{Rizzo:2014xma}.

\begin{figure}[t]
\centering
\includegraphics[clip,width=0.45\linewidth]{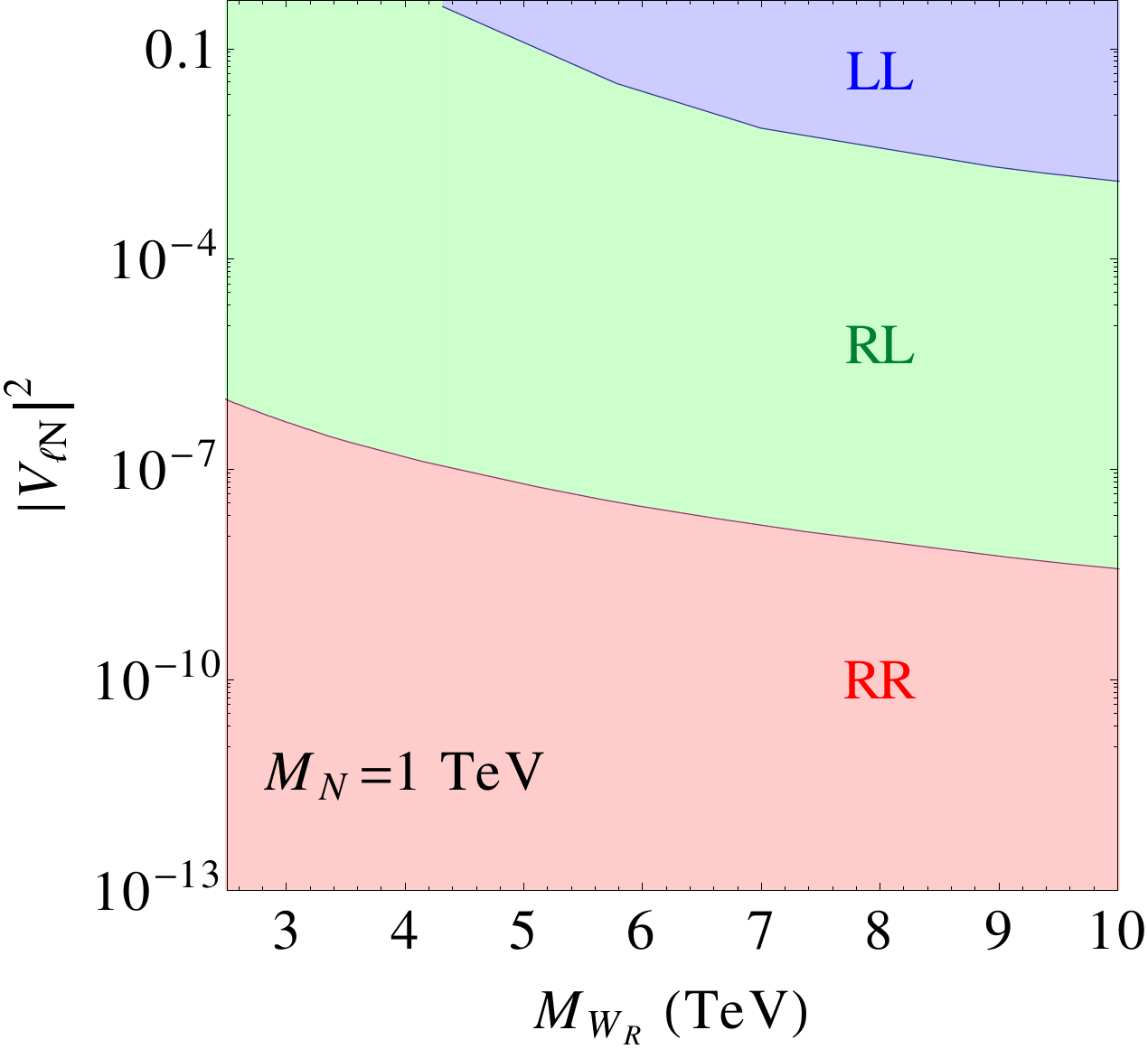} 
\caption{Phase diagram illustrating the dominance of different channels, namely, LL, RL and RR as shown in Figure~\ref{fig:LR} (left panel) in different regions of LRSM parameter space. Herer we have chosen $M_N=100$ GeV for illustration. The gray shaded region labeled `Seesaw' is the scale of mixing parameter $|V_{\ell N}|^2$ as expected in canonical seesaw [cf.~\eqref{Mnu}]. The brown shaded region labeled `EWPD' is the typical range of mixing ruled out from electroweak precision data (see Section~\ref{sec:ewpd}).}
\label{fig:col-LR}
\end{figure}

In variations of low-scale LRSM with large left-right neutrino mixing~\cite{Dev:2013oxa}, there will be new contribution to the like-sign dilepton signal due to mixed RH-LH currents~\cite{Chen:2013foz, Dev:2013vba}, in addition to the purely RH and LH contributions. Note that the amplitude for the RR diagram in Figure~\ref{fig:LR}~(left) is independent of $V_{lN}$, and hence, does not probe the full seesaw matrix \eqref{eq:matr}. On the other hand, the RL diagram is sensitive to the heavy-light mixing~\cite{Chen:2013foz}, and in fact, the dominant channel over a fairly large range of model parameter space, as illustrated in Figure~\ref{fig:col-LR}. Thus, a combination of the diagrams shown in Figure~\ref{fig:LR}~(left) is essential to fully explore the seesaw mechanism at the LHC. In this context, it is also useful to distinguish the RH gauge boson contributions to the collider signatures from the LH ones using different kinematic variables~\cite{Han:2012vk, Chen:2013foz}.

Recently, the CMS collaboration analyzed the $\sqrt s=8$ TeV LHC data with 19.7fb$^{-1}$ resulting in the most stringent direct bounds on the RH gauge boson masses up to $M_{W_R} = 3$~TeV~\cite{Khachatryan:2014dka}. While the analysis finds no significant departure from SM expectations, the LHC data exhibit an intriguing excess in $ee$ production with a local significance of $2.8\sigma$ for a candidate $W_R$ mass of $M_{W_R} \approx 2.1$~TeV; no excess is observed in the $\mu\mu$ channel. The excess could be interpreted as a hint for $W_R$ production, but with a smaller RH gauge coupling $g_R / g_L \simeq 0.6$~\cite{Deppisch:2014qpa, Heikinheimo:2014tba, Aguilar-Saavedra:2014ola, Deppisch:2014zta}. If the excess turns out to be statistically significant in future with more data and independent scrutiny from ATLAS, it might be an evidence for L-R symmetry with high-scale parity breaking~\cite{Chang:1983fu}.

\section{Lepton Number Violation at the LHC and Leptogenesis}\label{sec:5}

The observed baryon asymmetry of the Universe~\cite{Planck:2015xua} is far too large compared to the SM expectation. This is because in the SM, the necessary $C\!P$ violation is too small and no first order phase transition can take place for the observed value of the Higgs mass. The mechanism of baryogenesis in the Universe thus requires physics beyond the SM. A large number of possible mechanisms to generate the observed baryon asymmetry have been proposed in the literature. A particularly interesting scenario that also connects to the neutrino mass mechanism is {\em leptogenesis}~\cite{fukugita:1986hr}. In its original formulation, the out-of-equilibrium decay of the same heavy Majorana neutrinos responsible for the Type-I seesaw mechanism would create a lepton asymmetry, which is then reprocessed into a baryon asymmetry through $(B+L)$-violating EW sphaleron transitions~\cite{kuzmin:1985mm} at or above the scale of EWSB. In `vanilla' leptogenesis with hierarchical heavy neutrino masses~\cite{buchmuller:2004nz}, the neutrino oscillation data impose a {\em lower} bound of $M_N \gtrsim 5\times 10^8$~GeV on the lightest RH neutrino mass~\cite{davidson:2002qv, buchmuller:2002rq, giudice:2003jh}, which is inaccessible to foreseeable laboratory experiments. Moreover, these high-scale thermal leptogenesis scenarios are in conflict with the upper limits on RH neutrino masses from EW naturalness arguments~\cite{Vissani:1997ys, casas:2004gh,  Farina:2013mla, Clarke:2015gwa}, and in many supergravity models, from the gravitino overproduction bound~\cite{khlopov:1984pf, ellis:1984eq, Ellis:1984er, Kawasaki:1994af, Cyburt:2002uv, kawasaki:2004qu, Kawasaki:2008qe}. These problems can be naturally avoided in the framework of resonant leptogenesis~\cite{Pilaftsis:1997dr, Pilaftsis:1997jf, pilaftsis:2003gt}, where the heavy-neutrino self-energy effects on the leptonic $C\!P$ asymmetry become dominant~\cite{Flanz:1994yx, Covi:1996wh} and get resonantly enhanced, even up to order 1~\cite{Pilaftsis:1997dr, Pilaftsis:1997jf}, when two heavy Majorana neutrinos have a small mass difference comparable to their decay widths. This allows successful thermal leptogenesis with low seesaw scale accessible to laboratory experiments, while maintaining agreement with other theoretical and experimental constraints~\cite{Pilaftsis:2004xx, Pilaftsis:2005rv, Deppisch:2010fr, Dev:2014laa, Dev:2014tpa}.

\begin{figure}[t]
\centering
\includegraphics[clip,width=0.47\linewidth]{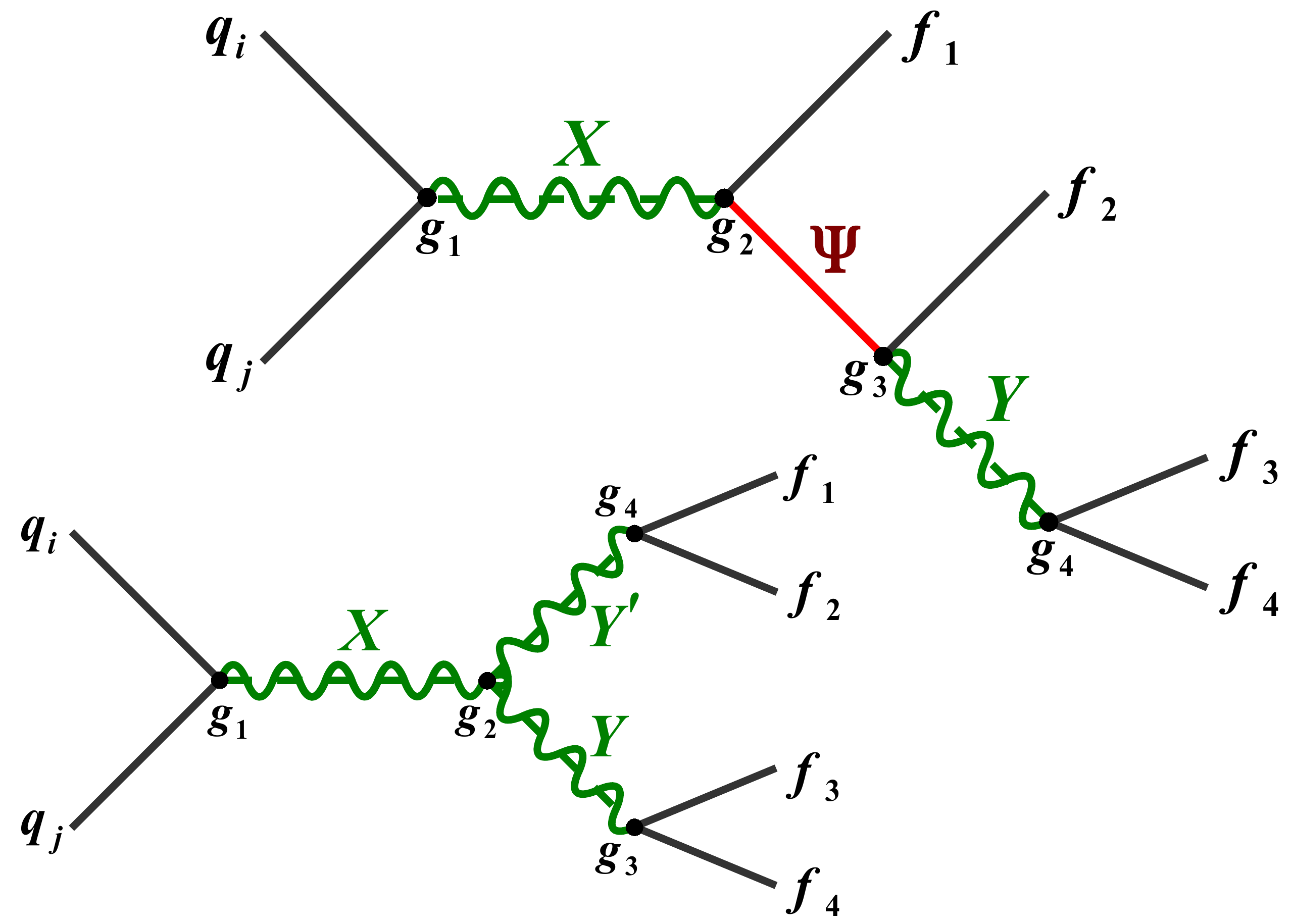}
\includegraphics[clip,width=0.49\linewidth]{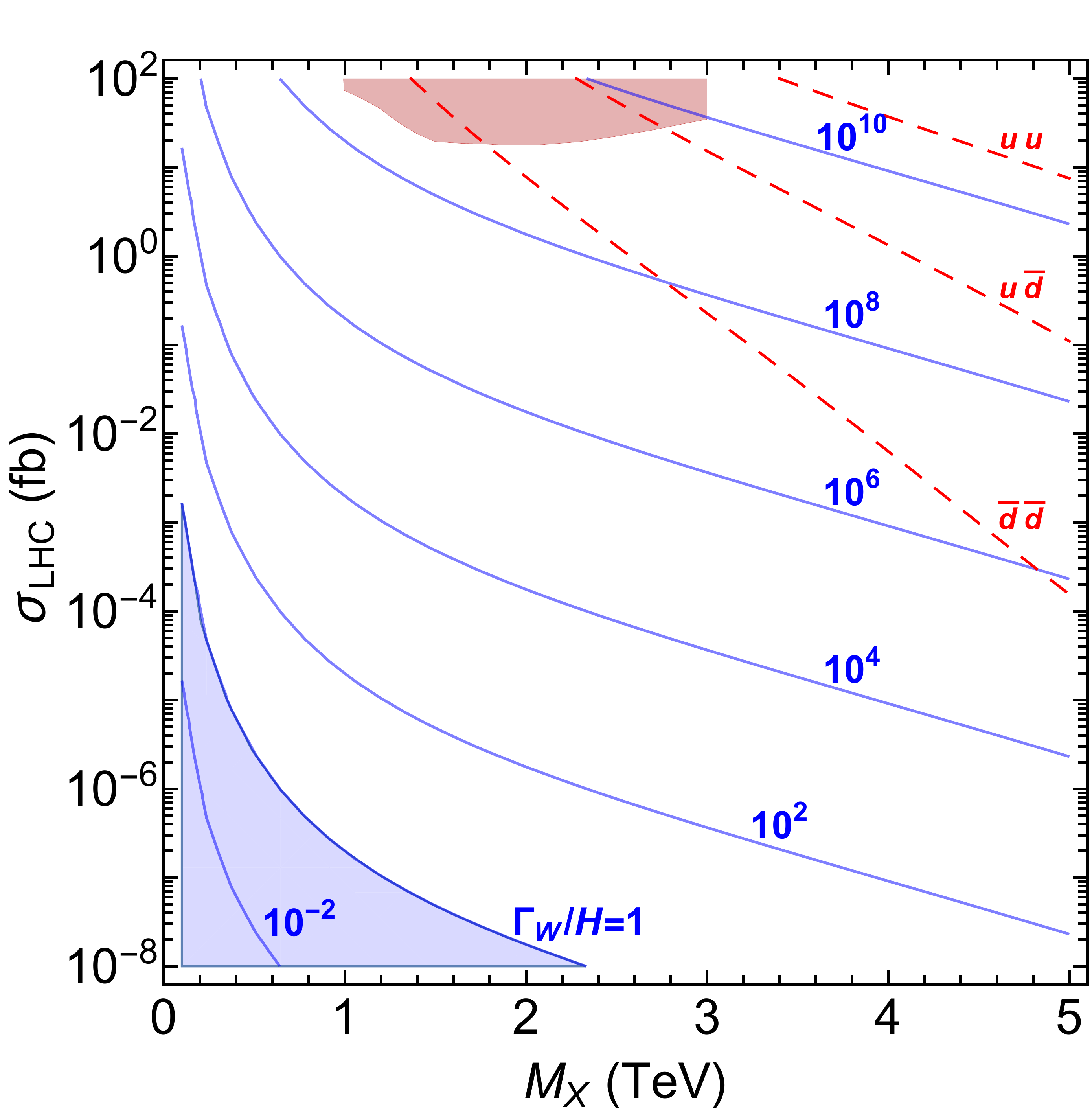}
\caption{{\em Left:} Generic diagrams for resonant same sign dilepton production $pp \to l^\pm l^\pm q q$ at the LHC. The intermediate particles are generic scalar or vector bosons $X$, $Y^{(')}$ and a fermion $\Psi$. Any two of the final state fermions $f_i$ can be leptons, depending on the transformation properties of the intermediate particles under the SM gauge group. 
{\em Right:} Induced lepton number washout rate $\Gamma_W/H$ at $T = M_X$ as a function of the LNV scale $M_X$ and the LHC cross section $\sigma_\text{LHC}$ (solid blue contours). The red dashed curves denote typically expected cross sections for gauge strength interactions. The shaded area is excluded by current LHC searches~\cite{Khachatryan:2014dka}.
}
\label{fig:decompositions}
\end{figure}

In addition, the observation of LNV at the LHC would have important consequences on the viability of general leptogenesis models. The issue of probing leptogenesis at the LHC has been studied in the context of the LRSM~\cite{Frere:2008ct, Dev:2014iva, Dhuria:2015wwa} but it is also possible to falsify a large class of high-scale leptogenesis scenarios if LNV was observed at the LHC~\cite{Deppisch:2013jxa, Deppisch:2014hva}. The analysis of~\cite{Deppisch:2013jxa} focuses on the resonant LNV process $pp \to l^\pm l^\pm jj$ involving generic intermediate particles as shown in Figure~\ref{fig:decompositions}~(left). Such a class of diagrams is generated by general decompositions of the corresponding 9-dimensional short range $0\nu\beta\beta$ operator~\cite{Helo:2013ika}. As an example, the specific realization in the LRSM is shown in Figure~\ref{fig:LR}~(left). The minimal lepton asymmetry washout rate $\Gamma_W$ induced by the process in Figure~\ref{fig:decompositions}~(left) is then related to the corresponding LHC cross section $\sigma_\text{LHC}$ as~\cite{Deppisch:2013jxa} 
\begin{align}
\label{eq:washout_factor_estimation}
  \log_{10}\frac{\Gamma_W}{H} &\gtrsim
  6.9 + 0.6\left( \frac{M_X}{\text{TeV}} - 1 \right) +
  \log_{10}\frac{\sigma_\text{LHC}}{\text{fb}},
\end{align}
where $H$ is the Hubble parameter at the scale $M_X$, i.e. the mass of the resonance in Figure~\ref{fig:decompositions}~(left). If $\Gamma_W/H \gg 1$, the dilution of a primordial net lepton number density, understood to be produced by a leptogenesis mechanism at a higher scale, is highly effective and the lepton asymmetry would be washed out before it can be converted by sphaleron processes. This result is illustrated graphically in Figure~\ref{fig:decompositions}~(right). Both \eqref{eq:washout_factor_estimation} and Figure~\ref{fig:decompositions}~(right) demonstrate that the observation of LNV at the LHC necessitates a very large lepton asymmetry washout. It would therefore rule out or strongly constrain leptogenesis scenarios above the scale $M_X$. Low scale scenarios, such as resonant leptogenesis discussed above where the lepton asymmetry is generated at scales lower than $M_X$, would not necessarily be constrained.

The approach is not limited to resonant same sign dilepton production but can be applied to any other process with $\Delta L \neq 0$, $\Delta (B-L) \neq 0$ and no missing energy at the LHC, and potentially at other future colliders. One example would be resonant pair-production of heavy particles, e.g. leptoquarks decaying to same-sign dilepton plus two jets~\cite{Kohda:2012sr} or singlet fermions decaying into a six fermion final state [cf. Figure~\ref{fig:zprime_diagram}]. Searches for high-energy LNV processes can therefore have a significant impact on models of leptogenesis, and baryogenesis in general.

\section{Supersymmetric Models} \label{sec:6}

In spite of the null results for SUSY searches so far at the LHC, SUSY still remains one of the most popular BSM scenarios due to its many attractive features, such as the gauge coupling unification, radiative EWSB, amelioration of the gauge hierarchy problem, natural DM candidates, connection to gravity, and so on (for a review, see e.g.~\cite{Haber:1984rc}). However, the simplest supersymmetric extension of the SM, namely, the Minimal Supersymmetric SM (MSSM) with conserved $R$-parity, does not accommodate non-zero neutrino masses. There are two common approaches for breaking the $L$-symmetry of MSSM to incorporate neutrino masses, and we briefly discuss their collider signatures in the following subsections. The possibility of connecting neutrino masses to SUSY-breaking sector has also been investigated (see e.g.~\cite{Borzumati:1999sp, Frere:1999uv, ArkaniHamed:2000bq, Demir:2007dt, Figueiredo:2014gpa}).
\subsection{Supersymmetric Seesaw}\label{sec:6.1}
Just as light neutrino masses can be generated in a seesaw extension of the SM, a supersymmetric generalization of the seesaw mechanism can accommodate massive neutrinos in SUSY models with conserved $R$-parity.  The superpotential of the Type-I seesaw extension of the MSSM is given by 
\begin{align}
{\cal W}  \ = \ & {\cal W}_{\rm MSSM}+ h_{ij} \epsilon_{ab}\widehat{L}_i^a \widehat{H}^b_u\widehat{N}_j^c+\frac{1}{2}(M_N)_{ij}\widehat{N}_i^c\widehat{N}_j^c \; ,
\label{sup1}
\end{align}
where $\widehat{L}_i$ represents the chiral multiplet containing a $SU(2)_L$ lepton doublet $(\nu, e)_{Li}$ and its corresponding superpartner, $\widehat{H}_u$ represents the $Y=1$ Higgs doublet and its Higgsino superpartner, $\widehat{N}_i^c$ is a RH neutrino superfield, $i,j$ are family indices, $a,b$ are $SU(2)$ indices and $\epsilon_{ab}$ is the antisymmetric $SU(2)$ tensor.After EWSB, the neutrino mass matrix is given by~\eqref{eq:seesaw}, as in the non-SUSY case. Moreover, the SUSY analogue of the Majorana mass term in the sneutrino sector leads to sneutrino-antisneutrino mixing, which could give rise to a same-sign dilepton signal at colliders~\cite{Grossman:1997is, Hall:1997ah, Choi:2001fka, Honkavaara:2005fn, Dedes:2007ef}. The neutrino Yukawa couplings lead to additional LFV effects in slepton masses, $\mu\to e\gamma$ in particular, through renormalization group effects in high-scale seesaw models~\cite{Borzumati:1986qx, Hisano:1995nq, Hisano:1996cp, Casas:2001sr, Ellis:2002fe, Masiero:2004js,  Arganda:2005ji}. In low-scale SUSY seesaw models, new sources of LFV are present due to large neutrino Yukawa couplings and threshold effects from low-scale RH neutrinos and sneutrinos~\cite{Deppisch:2003wt, Deppisch:2004fa, Ilakovac:2012sh, Abada:2014kba}. In such scenarios, the LFV rates of $\mu-e$ conversion and $\mu\to 3e$ could be sizable, even bigger than the $\mu\to e\gamma$ rate. 

In low-scale SUSY seesaw models, the lightest superpartner of the RH neutrino with a small admixture of its LH counterpart could be another viable candidate for DM, if it happens to be the lightest supersymmetric particle (LSP)~\cite{ArkaniHamed:2000bq, Garbrecht:2006az, arina:2008bb, cerdeno:2008ep, Deppisch:2008bp, An:2011uq}. The sneutrino LSP scenario leads to distinct collider signatures,  such as long missing transverse energy tail and enhanced same-sign dilepton signal in gluino and squark cascade decays~\cite{Belanger:2011ny, BhupalDev:2012ru, Banerjee:2013fga, Arina:2013zca, Guo:2013asa}, unlike the neutralino LSP in the MSSM scenario.

\subsection{$R$-Parity Violation}\label{sec:6.2}
Within the SM, the requirement of gauge invariance automatically guarantees $B$ and $L$ conservation for all renormalizable interactions. However, this is not the case in the general MSSM, and the following $B$ and $L$ violating terms are allowed in the superpotential: 
\begin{align}
{\cal W}_{\rm RPV} \ = \ \varepsilon_i \epsilon_{ab} \widehat{L}_i^a \widehat{H}_u^b + \lambda_{ijk} \epsilon_{ab}\widehat{L}_i^a \widehat{L}_j^b \widehat{E}_k^c + \lambda'_{ijk} \epsilon_{ab} \widehat{L}_i^a \widehat{Q}_j^b \widehat{D}_k^c + \lambda''_{ijk} \epsilon_{lmn} \widehat{U}_i^l \widehat{D}_j^{cm} \widehat{D}_k^{cn} , 
\label{rpv}
\end{align}
where $i,j,k$ are family indices, $a,b$ are $SU(2)$ indices, $l,m,n$ are color indices, and the chiral multiplets $\widehat{Q}, \widehat{U}^c, \widehat{D}^c, \widehat{E}^c$ respectively represent the $(u,d)_L, u^c_L, d^c_L, e^c_L$ and the corresponding superpartners.  Within the MSSM, these terms are forbidden by imposing an additional global symmetry that leads to the conservation of $R$-parity, given by $R=(-1)^{3(B-L)-2S}$ (with $S$ being the spin of the component field)~\cite{Farrar:1978xj}. The first term in~\eqref{rpv} leads to bilinear $R$-parity violation (BRPV), while the remaining three terms collectively give rise to trilinear $R$-parity violation (TRPV). 

The RPV models provide an alternative way to incorporate massive neutrinos with the minimal particle content of the MSSM (for a review, see e.g.~\cite{Barbier:2004ez}). The BRPV model is the simplest one~\cite{Hall:1983id, Joshipura:1994ib, Hempfling:1995wj,  Smirnov:1995ey, Nowakowski:1995dx, Nilles:1996ij, Diaz:1997xc, Joshipura:1998fn, Kaplan:1999ds, Hirsch:2000ef}, and has several interesting consequences that can be probed at collider experiments (see e.g.~\cite{hirsch:2004he, FileviezPerez:2012mj, Ghosh:2014ida}).  Since the distinction between the matter doublet superfields $\widehat{L}_i$ and the Higgs doublet superfields $\widehat{H}_d, \widehat{H}_u$ is lost in these models, it allows the mixing of neutrinos and neutralinos, sleptons and Higgs bosons, charged leptons and charginos. One linear combination of neutrino fields develops a Majorana mass at tree-level via mixing with Higgsinos, while the other combinations can acquire masses at loop-level via the trilinear couplings. Note that the trilinear coefficients $\lambda,\lambda'$ and $\lambda''$ are constrained from data on various low-energy $B$- and $L$-violating processes~\cite{Bhattacharyya:1997vv}. 

The collider phenomenology of RPV models has quite distinct features from that of the MSSM~\cite{Barbier:2004ez}. In particular, the LSP is unstable, and hence, not all SUSY decay chains lead to a large missing energy at colliders. The phenomenology of pair-produced SUSY particles is also modified due to new RPV decay chains. In addition, SUSY particles can now be singly produced, e.g. $s$-channel resonant production of sneutrinos in $e^+e^-$ collisions~\cite{Dimopoulos:1988jw, Kalinowski:1997bc, Erler:1996ww} and charged sleptons in hadron collisions~\cite{Dimopoulos:1988fr, Dreiner:2000vf, Dreiner:2012np}. RPV models also lead to sneutrino-antisneutrino mixing~\cite{Grossman:1998py} and other low energy effects, such as $0\nu\beta\beta$~\cite{Mohapatra:1986su, Babu:1995vh, Hirsch:1995zi}.    

\section{Conclusions and Outlook}\label{sec:7}
The discovery of neutrino oscillations has provided us with the first conclusive experimental evidence for the existence of new physics beyond the SM. Therefore, just as the postulate of the very existence of the neutrino led to the formulation of the theory of weak interactions, an essential ingredient for the stupendous success of the SM, a clear understanding of the neutrino mass mechanism could as well be the first beacon of physics beyond the SM. Therefore, it is very important to explore the experimental signatures of various neutrino mass models to pin down the underlying new physics. In this brief review, we have discussed some low-scale neutrino mass mechanisms  accessible to current and future experiments. In particular, we focused on the simplest Type-I seesaw model and summarized the current experimental constraints on the sterile neutrino mass and its mixing with active neutrinos. We have discussed the future discovery prospects of a heavy neutrino, within the minimal setup as well as involving extended gauge/Higgs sectors, with a particular emphasis on the energy frontier, in light of the upcoming run-II phase of the LHC and the proposed future colliders at both energy and intensity frontiers. A better picture of the neutrino portal might have far-reaching implications for the beyond SM scenarios in general, including the puzzles of matter-antimatter asymmetry and nature of dark matter in our Universe. In this context, we should emphasize the importance of complementary and synergetic explorations in the low-energy sector at the intensity frontier, as well as cosmological observations at the cosmic frontier, a combination of which is essential to fully unravel the mysteries of the neutrino world.        

\section*{Acknowledgments} 
P.S.B.D. would like to thank Oliver Fischer and Elena Graverini for sharing their data files on EWPD and SHiP limits respectively, and Marco Drewes for helpful comments on the draft. The work of F.F.D. is supported by the STFC grants ST/J000515/1 and ST/G000484/1, and the London Centre for Terauniverse Studies (LCTS), using funding from the European Research Council via the Advanced Investigator Grant 267352. 
The work of P.S.B.D. and A.P. is supported by the Lancaster-Manchester-Sheffield Consortium for Fundamental Physics under STFC grant ST/L000520/1. 

\section*{References}
\bibliographystyle{h-physrev}
\bibliography{NeutrinosLHCv3}
\end{document}